\documentclass[pdflatex,bst/sn-chicago]{sn-jnl}

\theoremstyle{thmstyleone}%
\theoremstyle{thmstyletwo}%

\theoremstyle{thmstylethree}%

\usepackage[utf8]{inputenc}
\usepackage[T1]{fontenc}

\usepackage{amsmath,amssymb,amsfonts,amsthm,mathrsfs,nicefrac,bm}

\usepackage[table,dvipsnames]{xcolor}
\usepackage{graphicx}
\usepackage{subcaption}
\usepackage[percent]{overpic}
\usepackage{wrapfig}

\usepackage{array,booktabs,colortbl,makecell,multirow,siunitx,tabularx,adjustbox}

\usepackage{algorithm,algpseudocode,algorithmicx,listings}

\usepackage[numbers]{natbib}
\usepackage{url}
\usepackage{hyperref}
\usepackage[capitalise]{cleveref}
\usepackage{doi}

\usepackage{xspace,textcomp,pifont,manyfoot,placeins,float,setspace,newfloat,enumitem,verbatim,microtype}
\usepackage{arydshln}

\usepackage[nolist,nohyperlinks]{acronym}  

\usepackage{newfloat}
\renewcommand{\figurename}{Fig.}

\DeclareFloatingEnvironment[
    fileext=loext,
    name=Extended Data Fig.
]{extendeddatafigure}

\DeclareCaptionLabelSeparator{pipe}{ \textbar{} }

\DeclareFloatingEnvironment[
    name=Extended Data Table
]{extendeddatatable}

\DeclareFloatingEnvironment[
    name=Supplementary Data Table,
]{supplementarydatatable}

\DeclareFloatingEnvironment[
    name=Supplementary Data Fig.,
]{supplementarydatafigure}

\begin{document}

\newcommand{\medimage}{MedImageInsight\xspace}
\newcommand{\panderm}{PanDerm\xspace}
\newcommand{\biomedclip}{BiomedCLIP\xspace}
\newcommand{\clip}{CLIP\xspace}
\newcommand{\lvm}{LVM-Med\xspace}
\newcommand{\resnet}{ResNet\xspace}
\newcommand{\siglip}{SigLIP\xspace}
\newcommand{\vit}{VIT-Base\xspace}
\newcommand{\derm}{DermINO\xspace}
\newcommand{\dino}{DINOv2\xspace}
\newcommand{\mae}{MAE\xspace}
\newcommand{\moco}{MoCov3\xspace}
\newcommand{\swav}{SWAV\xspace}
\newcommand{\bio}{BioMistral\xspace}
\newcommand{\coder}{CODER\xspace}


\newcommand{\hospital}{CSID-CJFH.\xspace}
\newcommand{\web}{LESION 130ks.\xspace}

\newcommand{\classone}{GLD6\xspace}
\newcommand{\classtwo}{MPL5\xspace}
\newcommand{\classthree}{SKFeat4\xspace}
\newcommand{\classfive}{SID2\xspace}
\newcommand{\classsix}{BNFeat6\xspace}
\newcommand{\classseven}{NNFeat6\xspace}
\newcommand{\classeight}{VWCD4\xspace}
\newcommand{\classnine}{DFFeat6\xspace}
\newcommand{\classten}{HFeat10\xspace}
\newcommand{\onesevenkthree}{Fitzpatrick17k-3\xspace}
\newcommand{\onesevenknine}{Fitzpatrick17k-9\xspace}
\newcommand{\dermsevenpt}{Derm7pt\xspace}
\newcommand{\pad}{PAD-UFES-20\xspace}

\newcommand{\classfourone}{MSD2\xspace}
\newcommand{\classfourtwo}{MTD2\xspace}
\newcommand{\acnezerofour}{ACNE04\xspace}

\newcommand{\ddi}{DDI\xspace}
\newcommand{\onesevenktwo}{Fitzpatrick17k-2\xspace}
\newcommand{\mednode}{MED-Node\xspace}
\newcommand{\phtwo}{PH2(cls)\xspace}

\newcommand{\skincancer}{Skincancer\xspace}
\newcommand{\ISICsix}{ISIC2016\xspace}
\newcommand{\ISICseven}{ISIC2017\xspace}
\newcommand{\ISICeight}{ISIC2018\xspace}
\newcommand{\segmentphtwo}{PH2(seg)\xspace}

\newcommand{\skincap}{SkinCAP\xspace}


\newcommand{\biovil}{BioViL-T\xspace}
\newcommand{\raddino}{R\textsc{ad}-DINO\xspace}




    \title[DermINO]{DermINO: Hybrid Pretraining for a Versatile Dermatology Foundation Model}


\author[1,4]{\fnm{Jingkai} \sur{Xu}}
\equalcont{These authors contributed equally to this work.}

\author*[2]{\fnm{De} \sur{Cheng}}\email{dcheng@xidian.edu.cn}
\equalcont{These authors contributed equally to this work.}

\author[2]{\fnm{Xiangqian} \sur{Zhao}}
\equalcont{These authors contributed equally to this work.}

\author[1,5]{\fnm{Jungang} \sur{Yang}}
\equalcont{These authors contributed equally to this work.}

\author[3]{\fnm{Zilong} \sur{Wang}}
\equalcont{These authors contributed equally to this work.}

\author[3]{\fnm{Xinyang} \sur{Jiang}}
\equalcont{These authors contributed equally to this work.}

\author[3]{\fnm{Xufang} \sur{Luo}}

\author[1,6]{\fnm{Lili} \sur{Chen}}

\author[1,7]{\fnm{Xiaoli} \sur{Ning}}

\author[1]{\fnm{Chengxu} \sur{Li}}

\author[1,5]{\fnm{Xinzhu} \sur{Zhou}}

\author[1,4]{\fnm{Xuejiao} \sur{Song}}

\author[1,8]{\fnm{Ang} \sur{Li}}

\author[1,8]{\fnm{Qingyue} \sur{Xia}}

\author[1,5]{\fnm{Zhou} \sur{Zhuang}}

\author[1,8]{\fnm{Hongfei} \sur{Ouyang}}

\author[1]{\fnm{Ke} \sur{Xue}}

\author[1]{\fnm{Yujun} \sur{Sheng}}

\author[1]{\fnm{Rusong} \sur{Meng}}

\author[11]{\fnm{Feng} \sur{Xu}}

\author[2]{\fnm{Xi} \sur{Yang}}

\author[10]{\fnm{Weimin} \sur{Ma}}

\author[10]{\fnm{Yusheng} \sur{Lee}}

\author[3]{\fnm{Dongsheng} \sur{Li}}

\author[2]{\fnm{Xinbo} \sur{Gao}}

\author[9]{\fnm{Jianming} \sur{Liang}}

\author*[3]{\fnm{Lili} \sur{Qiu}}\email{liliqiu@microsoft.com}

\author*[2]{\fnm{Nannan} \sur{Wang}}\email{nnwang@xidian.edu.cn}

\author*[1,4,6]{\fnm{Xianbo} \sur{Zuo}}\email{zuoxianbo@qq.com}

\author*[1,5,6,7,8]{\fnm{Cui} \sur{Yong}}\email{wuhucuiyong@vip.163.com}

\affil[1]{\orgdiv{Department of Dermatology}, \orgname{China-Japan Friendship Hospital}, \city{Beijing}, \postcode{100029}, \country{China}}

\affil[2]{\orgdiv{State Key Laboratory of Integrated Services Networks (ISN)}, \orgname{Xidian University}, \city{Shaanxi}, \postcode{710071}, \country{China}}

\affil[3]{\orgname{Microsoft Research Asia}, \city{Shanghai}, \postcode{200232}, \country{China}}

\affil[4]{\orgdiv{Big Data Center}, \orgname{China-Japan Friendship Hospital}, \city{Beijing}, \postcode{100029}, \country{China}}

\affil[5]{\orgname{Peking University China-Japan Friendship School of Clinical Medicine}, \city{Beijing}, \postcode{100029}, \country{China}}

\affil[6]{\orgdiv{Institute of Clinical Medicine for China-Japan Friendship Hospital}, \orgname{China-Japan Friendship Hospital}, \city{Beijing}, \postcode{100029}, \country{China}}

\affil[7]{\orgdiv{Department of Dermatology}, \orgname{China-Japan Friendship Hospital, Capital Medical University}, \city{Beijing}, \postcode{100029}, \country{China}}

\affil[8]{\orgdiv{China-Japan Friendship Hospital (Institute of Clinical Medical Sciences)}, \orgname{Chinese Academy of Medical Sciences \& Peking Union Medical College}, \city{Beijing}, \postcode{100029}, \country{China}}

\affil[9]{\orgdiv{Biomedical Informatics and Data Science}, \orgname{Arizona State University}, \city{Phoenix}, \state{AZ} \postcode{85281}, \country{USA}}

\affil[10]{\orgname{Scientific-skincare Innovation Alliance (SIA)}, \city{Shanghai}, \postcode{201103}, \country{China}}

\affil[11]{\orgdiv{Department of Dermatology}, \orgname{Huashan Hospital, Fudan University}, \city{Shanghai}, \postcode{200040}, \country{China}}

\abstract{
Skin diseases impose a substantial burden on global healthcare systems, driven by their high prevalence (affecting up to 70\% of the population), complex diagnostic processes, and a critical shortage of dermatologists  in resource-limited areas.  While artificial intelligence(AI) tools have demonstrated promise in dermatological image analysis, current models face limitations—they often rely on large, manually labeled datasets and are built for narrow, specific tasks, making them less effective in real-world settings. 
To tackle these limitations, we present \derm, a versatile foundation model for dermatology. Trained on a curated dataset of 432,776 images from three sources (public repositories, web-sourced images, and proprietary collections), \derm incorporates a novel hybrid pretraining framework that augments the self-supervised learning paradigm through semi-supervised learning and knowledge-guided prototype initialization. 
This integrated method not only deepens the understanding of complex dermatological conditions, but also substantially enhances the generalization capability across various clinical tasks. 
Evaluated across 20 datasets, \derm consistently outperforms state-of-the-art models across a wide range of tasks. It excels in high-level clinical applications including malignancy classification, disease severity grading, multi-category diagnosis, and dermatological image caption, while also achieving state-of-the-art performance in low-level tasks such as skin lesion segmentation. Furthermore, \derm demonstrates strong robustness in privacy-preserving federated learning scenarios and across diverse skin types and sexes. 
In a blinded reader study with 23 dermatologists, \derm achieved 95.79\% diagnostic accuracy (versus clinicians' 73.66\%), and AI assistance improved clinician performance by 17.21\%.
These findings underscore DermINO's strong potential to enhance dermatology AI tools used in screening, diagnosis, and telemedicine applications.

}    




    \maketitle
    
    \section{Introduction}\label{sec1}

Skin diseases represent one of the leading global health burdens, affecting up to 70\% of the global population~\cite{globalburden2010,globalburden2021,primarycare1,primarycare2,primarycare3}.
Diagnosing skin diseases remains a significant challenge due to several reasons. 
First, their clinical presentations are highly heterogeneous, spanning a broad spectrum of conditions from benign inflammatory disorders and chronic infections to malignant tumors.
Second, there is considerable variability in lesion appearance, influenced by skin type, anatomical location, and disease progression.
Third, there is a global shortage of experienced dermatologists, especially in resource-limited regions.
As a result, misdiagnoses or delays in diagnosis are common, leading to increased patient morbidity and substantial healthcare burdens.
In this context, accurate and efficient image-based analysis has become a vital tool in supporting dermatological clinical decision-making.

Recent advances in deep learning have enabled AI systems to achieve expert-level performance in certain dermatological tasks~\cite{brinker2019deep,liu2020deep}. However, most current models rely heavily on large, manually annotated datasets, which are costly and difficult to scale. In addition, traditional AI approaches are often designed for narrow, specific tasks, limiting their ability to generalize across the wide spectrum of skin conditions encountered in practice.

As an emerging and transformative paradigm in AI, foundation models offer a promising solution to these challenges. 
By leveraging large-scale self-supervised pretraining on diverse datasets, they can learn rich and transferable feature representations, enabling adaptation to new tasks with minimal annotated data. Foundation models have already demonstrated remarkable generalization and multi-task capabilities in medical domains such as radiology~\cite{radio}, pathology~\cite{uni}, and ophthalmology~\cite{Retfound}, suggesting significant potential for transforming AI applications in dermatology.

Despite recent advances, dermatology has yet to fully benefit from foundation models, largely due to several domain-specific challenges. These include the exceptional visual diversity of skin conditions, the need for specialized clinical knowledge to interpret subtle features, and a persistent scarcity of high-quality, expertly labeled datasets.

Dermatological AI must support a wide spectrum of clinical tasks, ranging from high-level objectives such as malignancy classification, severity grading, multiclass diagnosis, and image captioning to low-level one like lesion segmentation. For example, malignancy classification can assist in the differentiation of malignant melanoma, while severity grading is important for conditions such as acne or melasma. Multi-class diagnosis enables the identification of various skin diseases in a single platform. Image captioning can provide automated descriptions of dermoscopic images to support documentation and communication. Low-level task such as lesion segmentation is fundamental for accurate measurement and localization of skin lesions. Together, these capabilities address key clinical needs and support comprehensive decision-making in dermatology. Moreover, dermatological AI must demonstrate robust performance in privacy-preserving settings, such as federated learning, while also maintaining a fair balance between accuracy and equity across different skin types and sexes.

\begin{figure*}[p]
    \centering
    \includegraphics[width=1\linewidth]{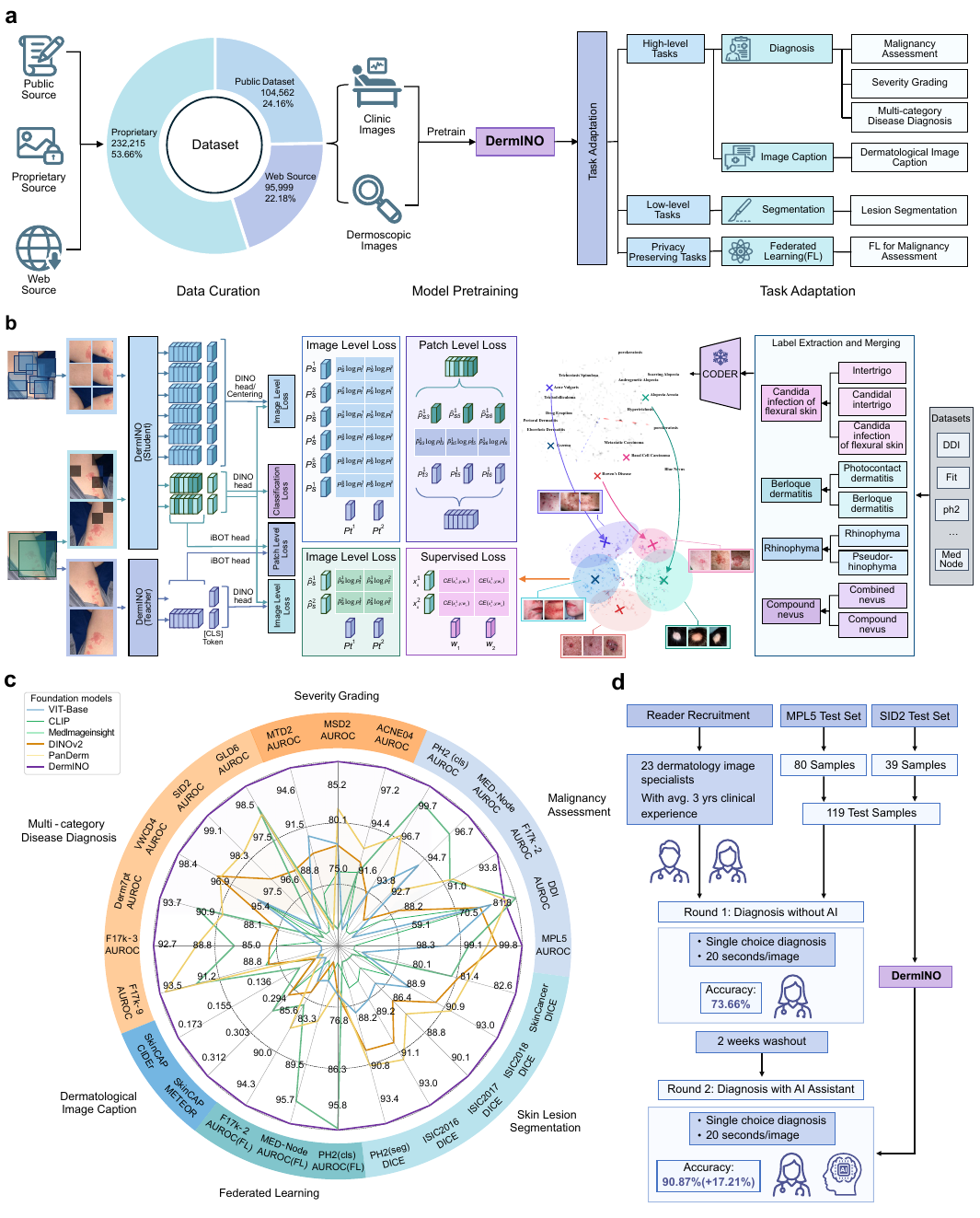} 
    \caption{\textbf{Overview of DermINO.}
    \textbf{a}, DermINO is pretrained on diverse dermatology datasets, including public, web-sourced, and proprietary images, and adapted for a wide range of high- and low-level tasks. It is also evaluated in a privacy-preserving federated learning scenario.
    \textbf{b}, DermINO uses a hybrid pretraining framework that combines self-supervised learning with supervised training on partially annotated data, and incorporates a knowledge-guided prototype initialization to enhance clinical relevance.
    \textbf{c}, DermINO outperforms state-of-the-art foundation models across 20 downstream datasets covering six dermatological tasks.
    \textbf{d}, A two-phase reader study with 23 dermatology specialists assessed DermINO's diagnostic accuracy and AI-assisted performance.
    }
    \label{fig:fig1_overview}
\end{figure*}

In response to these challenges, we introduce \derm, a large-scale versatile foundation model specifically designed for dermatological image analysis (Fig.\ref{fig:fig1_overview}). For pretraining, we curated a comprehensive dataset comprising 432,776 images, integrating public datasets, web-sourced images,  and proprietary clinical collections from our hospital. To address the scarcity of annotated data, we introduce a hybrid pretraining paradigm that enhances DINO self-supervised learning framework by combining large-scale self-supervised learning with supervised training on partially annotated data. To support a wide range of clinical tasks, we propose a domain knowledge-guided prototype initialization strategy, which encodes expert knowledge into prototypes using medical language models to facilitate learning of diverse and clinically meaningful semantics. Additionally, by incorporating patch-level loss, we further strengthen the model’s ability to capture fine-grained visual details, improving performance on both high-level and low-level tasks, such as clinical classification and skin lesion segmentation.

Compared to existing state-of-the-art general and medical foundation models, \derm demonstrates superior generalization across a broad spectrum of clinical tasks. These include malignancy assessment, severity grading, multi-category disease diagnosis, dermatological image captioning and segmentation, as validated on 20 datasets, all of which rely on comprehensive whole-image understanding. Furthermore, \derm achieves  significantly better performance on task that require pixel-level image analysis such as lesion segmentation. Through extensive reader studies with 23 dermatology specialists,  DermINO achieved 95.79\% diagnostic accuracy (versus clinicians' 73.66\%), and AI assistance also improved clinician performance by 17.21\%.
These results underscore the strong potential of this foundation model to empower dermatology AI systems for essential tasks such as screening, diagnosis, and grading, and to facilitate real-world application, such as teledermatology and smart medical device integration.

    \section{Results}\label{sec2}

To comprehensively evaluate the effectiveness of \derm, we conducted benchmark comparisons on both high-level clinical tasks and low-level image recognition tasks against three widely used categories of pretrained models.  
In the self-supervised vision models category, we included the general-purpose model \dino~\cite{oquab2023dinov2}, along with medical domain-specific models such as \lvm~\cite{lvmmed} and \panderm~\cite{panderm}. For vision-language models, our evaluation covered general models like \clip~\cite{clip} and \siglip~\cite{siglip}, as well as medical-specific counterparts including \biomedclip~\cite{biomedclip} and \medimage~\cite{medimage}. In the supervised vision models category, we assessed widely used general-purpose baselines such as \resnet~\cite{resnet} and \vit~\cite{vit_base}. The proposed \derm is a hybrid pretraining framework that integrates self-supervised learning with domain-guided supervision to leverage both general representation power and domain relevance. \derm is further evaluated through a reader study, in which its diagnostic accuracy is compared with that of dermatology image specialists using a representative set of dermatology images. The study also assesses the extent to which DermINO-assisted support improved dermatologists' performance.

\subsection{\derm generalizes to various high-level clinical tasks}

To assess the generalization capability of \derm in supporting a broad spectrum of high-level diagnostic tasks, we evaluated its performance across four clinically relevant dermatological applications, using a total of 14 test datasets. These tasks include malignancy assessment, severity grading, multi-category disease diagnosis, and dermatological image caption. 
The evaluation of the first three tasks (malignancy assessment, severity grading and multi-category disease diagnosis) were conducted under both image retrieval and classification settings. 
In the retrieval setting, image classification is achieved by retrieving the top-$k$ most similar images from a candidate pool based on feature representations extracted by the foundation model. The predicted label is then determined by majority voting among the labels of the retrieved images. In the classification setting, we append a linear classifier to the foundation model and fine-tune the model using labeled training data for each specific task.
The detailed implementation for the two experiment settings will be elaborated in the Methods  section. Two evaluation metrics are used for the first three tasks: macro-averaged area under the receiver operator characteristic (AUROC)~\cite{AUROC} and macro-averaged F1 score~\cite{F1}. 
For the image captioning task, evaluation was based on a comprehensive set of language metrics, including BLEU-1, BLEU-2~\cite{Bleu}, METEOR~\cite{meteor}, CIDEr~\cite{cider}, ROUGE-1, and ROUGE-L~\cite{rouge}.

\subsubsection{Malignancy assessment}

Accurate assessment of skin malignancies is crucial for effective skin cancer treatment, with direct implications for patient outcomes and healthcare resource allocation. To advance this objective, we curated a set of evaluation datasets to systematically compare the performance of various pretrained models in skin malignancy classification.

\begin{figure*}[h!]
    \centering
    \includegraphics[width=1\linewidth]{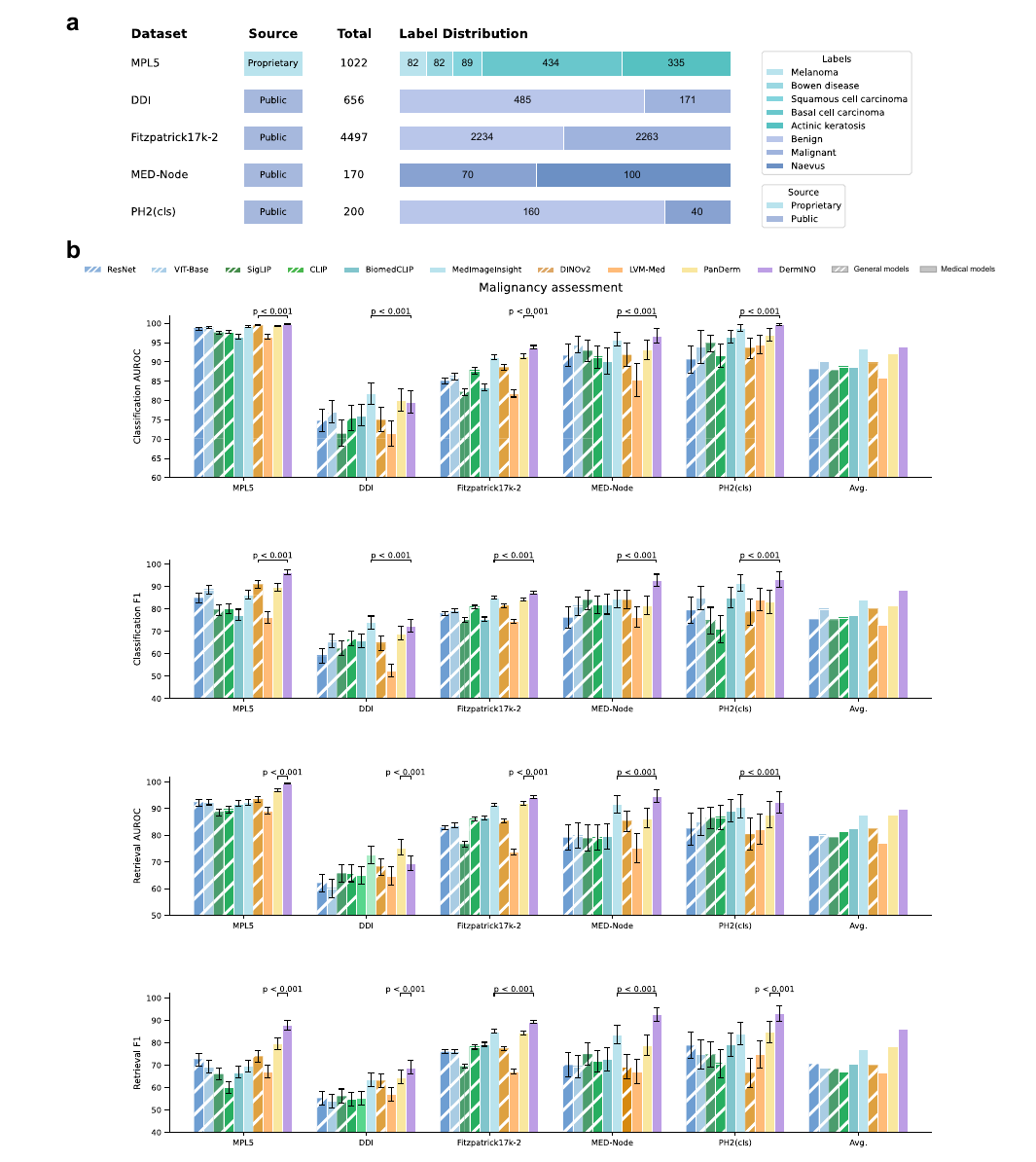}
    \caption{\textbf{Performance comparison on malignancy assessment.} 
    \textbf{a}, Overview of the five datasets used for malignancy assessment. 
    \textbf{b}, Comparison of \derm and nine existing foundation models across these five datasets in both the classification and retrieval settings. Performance is evaluated using AUROC and F1 score; error bars indicate one standard deviation.}
    \label{fig:clinic_1_combined}
\end{figure*}

We utilized four external datasets, \ddi (656 samples)\cite{ddi}, \onesevenktwo (4,497 samples)\cite{fit}, \mednode (170 samples)\cite{med_node}, and \phtwo (200 samples)\cite{ph2}, to construct binary malignancy classification tasks, differentiating malignant from benign skin conditions based on the provided labels. 
To further evaluate each model's ability in fine-grained differential diagnosis across multiple malignant conditions, we additionally incorporated an internal dataset, \classtwo (1,022 samples) (see Fig.\ref{fig:clinic_1_combined}), which includes five common types of malignant skin tumors: basal cell carcinoma, squamous cell carcinoma, malignant melanoma, Bowen's disease, and actinic keratosis.

\derm outperforms competing methods on both retrieval and classification task in terms of average AUROC and F1 scores, surpassing the next-best approach, \medimage \cite{medimage}, by 2.31\% in average retrieval AUROC, 0.53\% in average classification AUROC, and 4.03\% in average classification F1 score.  
We find that foundation models trained on dermatology-specific data, especially \panderm and \medimage, consistently outperform general-domain models in both retrieval and classification tasks. These results highlight the importance of incorporating domain-specific medical knowledge, which significantly enhances model performance in skin malignancy assessment.

\subsubsection{Severity grading}

\begin{figure*}[h!]
    \centering
    \includegraphics[width=1\linewidth]{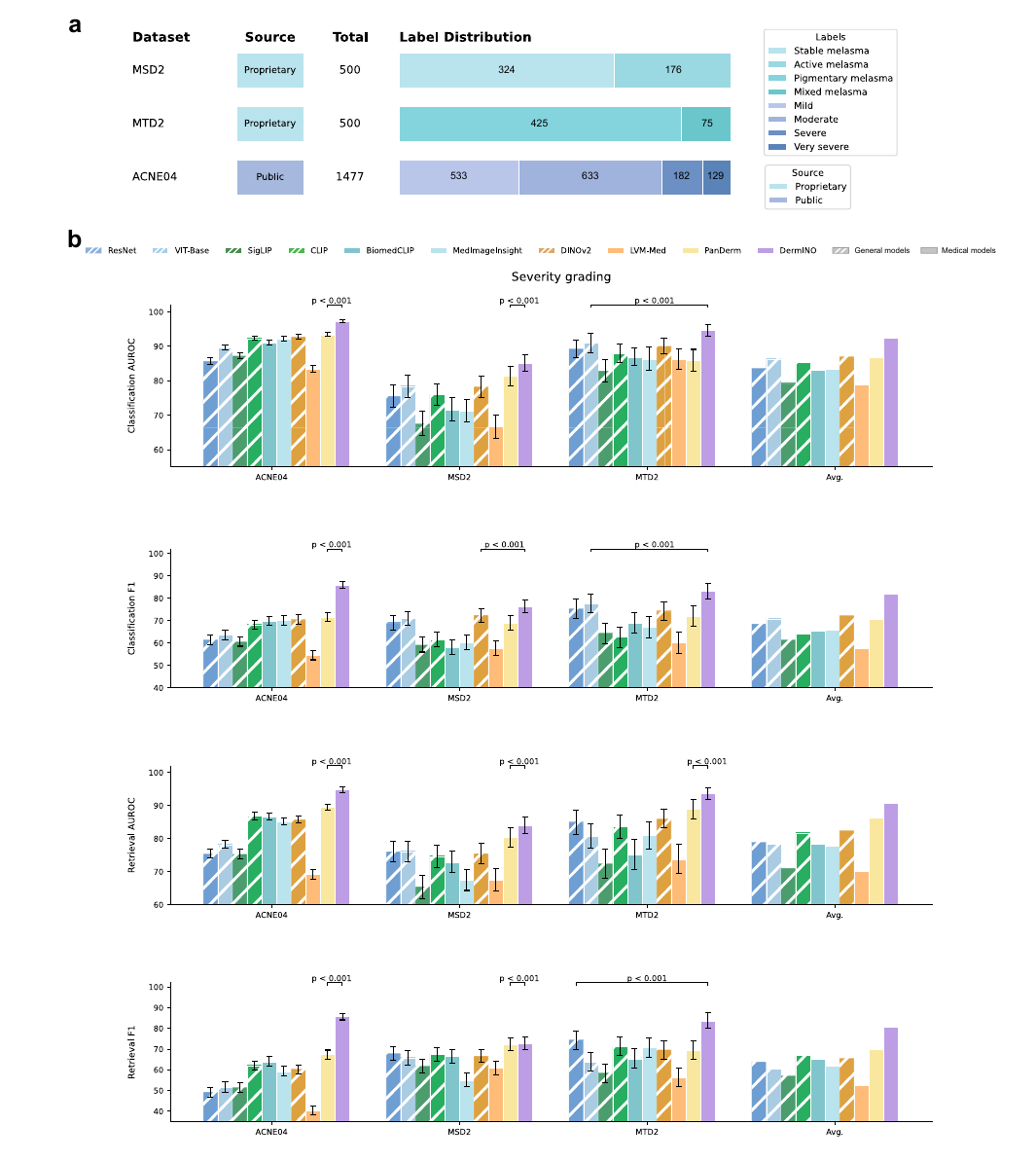}
    \caption{\textbf{Performance comparison on severity grading.} 
    \textbf{a}, Overview of the three datasets used for severity grading. 
    \textbf{b}, Comparison of \derm and nine existing foundation models across these three datasets in both the classification and retrieval settings. Performance is evaluated using AUROC and F1 score; error bars indicate one standard deviation.}
    \label{fig:clinic_2_combined}
\end{figure*}

In this section, we compare \derm with state-of-the-art foundation models on the task of dermatological disease severity grading, which is critical for clinical evaluation and therapeutic decision-making. Performance is assessed on three datasets: two internal datasets \classfourone (500 samples) and \classfourtwo (500 samples), and one widely adopted public benchmark, \acnezerofour (1,477 samples)~\cite{acne04} (see Fig.~\ref{fig:clinic_2_combined}). The internal datasets encompass both clinical typing and staging of melasma, while \acnezerofour provides four-level severity grading for acne vulgaris. These tasks demand fine-grained feature extraction and precise discrimination to capture subtle variations in disease severity. 


\derm achieves the highest performance across all three datasets (\(p < 0.001\)), outperforming the next-best method, \panderm\cite{panderm}, on severity grading by 4.46\% in average retrieval AUROC, and 11.01\% in average retrieval F1 score. Moreover, it exceeds \dino\cite{oquab2023dinov2} on classification by 5.31\% in average AUROC and 9.54\% in average F1 score.
Unlike in malignancy assessment, dermatology-specific pretrained models such as \medimage and \panderm do not show a clear advantage over general-domain foundation models on severity grading. Specifically, both \medimage and \panderm underperform compared to \dino in classification accuracy, and \medimage shows notably poor retrieval performance, trailing behind general-domain models like \dino and \clip. We attribute this to the greater complexity of severity grading, which requires highly detailed and fine-grained visual feature extraction. While \medimage and \panderm are pretrained on domain-specific data, their pretraining objectives (vision-language contrastive learning for \medimage and masked image reconstruction for \panderm) primarily focus on either global semantic alignment or low-level image reconstruction. Such objectives may not be comprehensive enough to capture the intricate visual cues necessary for severity grading. In contrast, \derm adopts a hybrid pretraining strategy that integrates both self-supervised and supervised signals, allowing it to learn richer and more task-relevant representations and thereby outperform both general-domain and medical-domain models.

\subsubsection{Multi-category disease diagnosis}

\begin{figure*}[h!]
    \centering
    \includegraphics[width=1\linewidth]{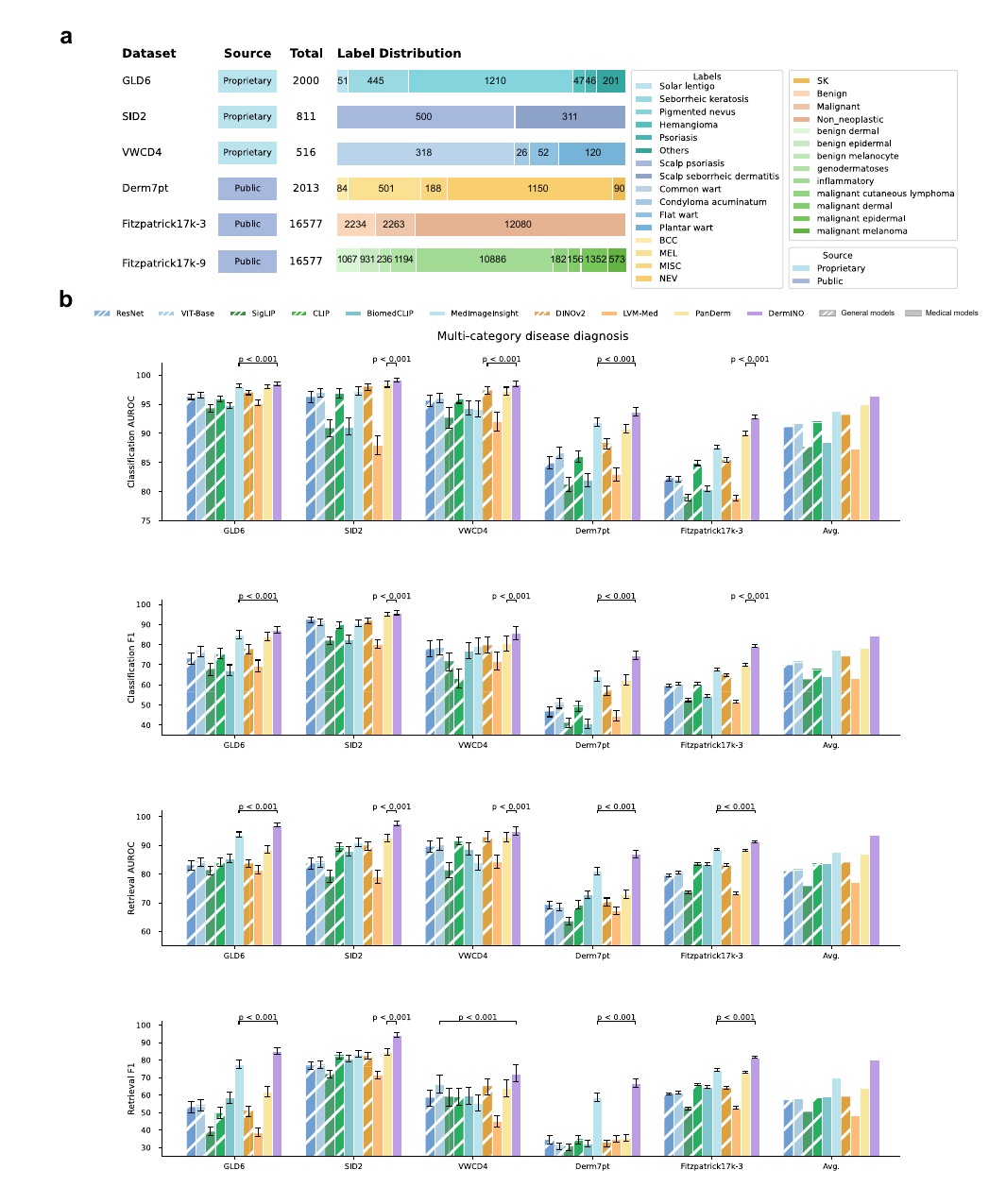}
    \caption{\textbf{Performance comparison on multi-category disease diagnosis.} 
    \textbf{a}, Overview of the six datasets used for multi-category disease diagnosis. 
    \textbf{b}, Comparison of \derm and nine existing foundation models across these six datasets in both the classification and retrieval settings. Performance is evaluated using AUROC and F1 score; error bars indicate one standard deviation.}
    \label{fig:clinic_3_combined}
\end{figure*}

In this section, we evaluate \derm’s generalization ability across a wide range of dermatological conditions. Model performance is assessed on three internal datasets: \classone (2,000 samples), which includes six disease categories distinguishing between skin tumors and inflammatory skin conditions, namely solar lentigo, seborrheic keratosis, pigmented nevus, hemangioma, psoriasis, and others; \classfive (811 samples), which focuses on differentiating scalp psoriasis from scalp seborrheic dermatitis; and \classeight (516 samples), comprising four types of viral warts. Additionally, we include three public datasets for evaluation across a more diverse spectrum of dermatological conditions: \onesevenkthree (16,577 samples)\cite{fit}, \onesevenknine (16,577 samples)\cite{fit} and \dermsevenpt (2,013 samples)\cite{derm7pt}, both covering a wide range of skin disease diagnoses (see Fig.~\ref{fig:clinic_3_combined}).

\derm outperforms all the comparing methods across six multi-category disease diagnosis datasets under both classification and retrieval setting (\(p<0.001\)). 
\derm outperforms the strongest competing model in the classification task, \panderm\cite{panderm}, by 1.38\% in average AUROC and 6.26\% in average F1 score . It also surpasses the leading model in the retrieval task, \medimage~\cite{medimage}, by 5.26\% in average AUROC and 10.04\% in average F1 score. 

Most foundation models pretrained on dermatology-specific data outperform their general-domain counterparts in multi-category disease diagnosis. Interestingly, although \panderm outperforms \medimage in the classification task, it performs relatively worse in retrieval. This discrepancy may stem from the different pretraining strategies adopted by the two models: \medimage is optimized for classification via vision-language alignment that emphasizes the learning of high-level semantic features, which is particularly beneficial for retrieval; in contrast, \panderm employs a reconstruction-based pretraining approach that focuses on preserving fine-grained visual details, thereby demonstrating stronger performance in classification.

\subsubsection{Dermatological image caption}

To further evaluate \derm’s ability to handle more complex and nuanced clinical scenarios, we extend the assessment beyond multi-category disease classification to include free-text descriptions that encompass a wide range of clinical observations. The image captioning task examines whether models can generate accurate and detailed captions that reflect both the visual characteristics and clinical relevance of dermatological images.


\derm significantly outperforms all baseline methods across all evaluation metrics on the publicly available dermatological image captioning dataset \skincap\cite{skincap}, with a statistically significant improvement  (\(p<0.001\)). \medimage ranks second, likely benefiting from its pretraining on medical-domain data paired with rich textual descriptions. In contrast, \panderm performs worse than the medical-domain model \biomedclip. This may be attributed to \biomedclip’s superior ability to align visual and textual modalities, stemming from its vision-language contrastive pretraining objective. By comparison, \panderm  which is trained using a pure self-supervised objective without any language supervision, lacks the necessary alignment capabilities for tasks requiring robust integration of visual and textual information.

\begin{figure*}[hp!]
    \centering
    \includegraphics[width=\textwidth]{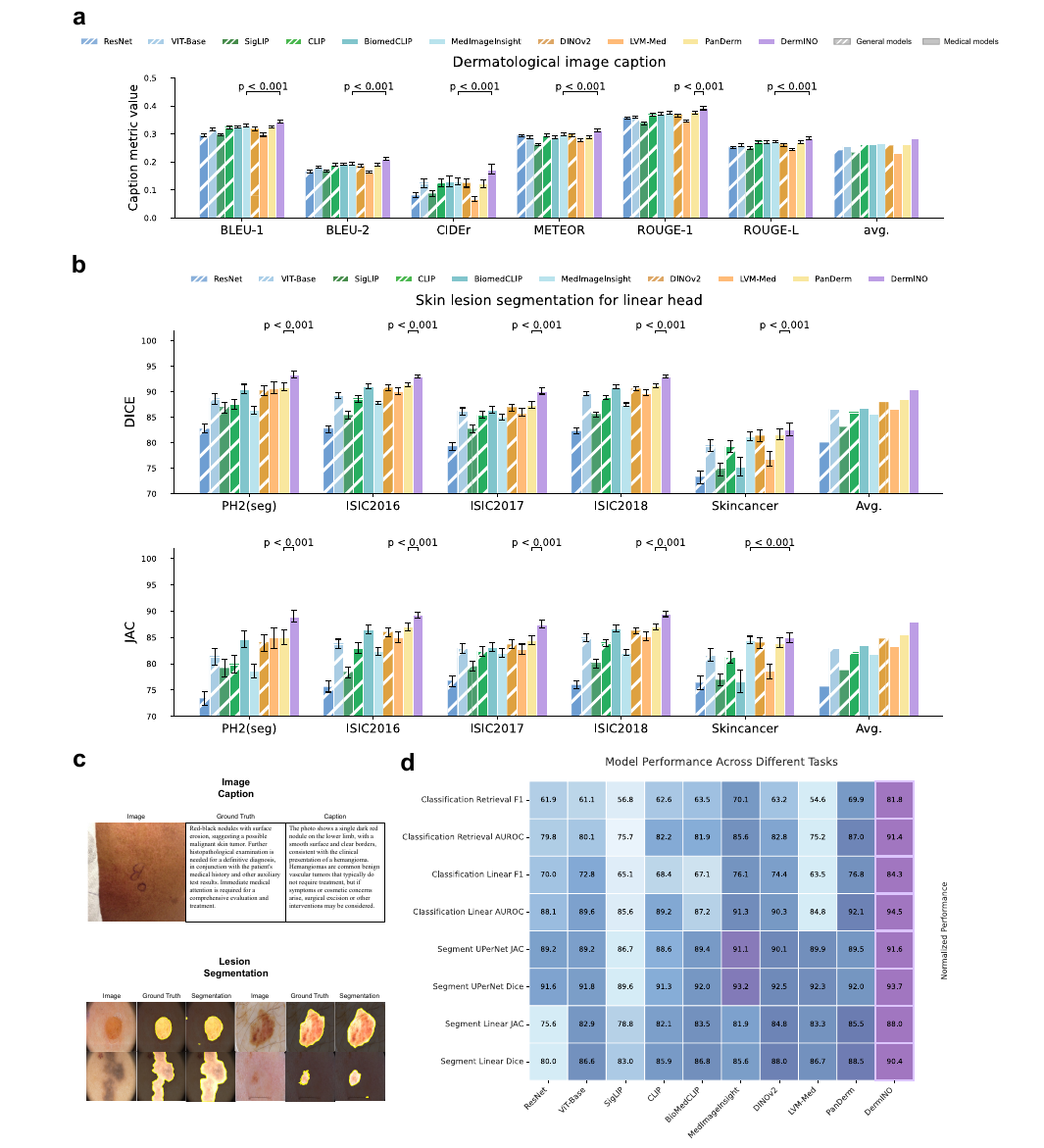}
    
    \caption{\textbf{DermINO performance on captioning, segmentation, and benchmarking across tasks.} \textbf{a}, Comparison of \derm with nine existing foundation models on a dermatological image captioning dataset, evaluated using multiple metrics to comprehensively assess caption quality. ~\textbf{b}, Performance comparison of \derm and nine foundation models on five lesion segmentation datasets using a linear segmentation head, evaluated by DICE coefficient and Jaccard Index (JAC). ~\textbf{c}, Visual examples illustrating the image captioning and segmentation tasks. ~\textbf{d}, Evaluation of performance balance across low-level and high-level dermatology tasks. The heatmap shows model performance on classification (retrieval and linear probing), segmentation with a linear head, and with UpperNet, highlighting the strengths and characteristics of each method across task types. }
    \label{fig:caption_segmentation_comparison}
\end{figure*}

\subsection{\derm improves low-level dermatology image recognition}

\subsubsection{Skin lesion segmentation}
Skin lesion segmentation plays a crucial role in clinical practice by enabling precise localization and delineation of skin abnormalities, which supports early diagnosis, treatment planning, and monitoring of skin diseases such as melanoma. 

We evaluate the models on five public lesion segmentation datasets: \segmentphtwo (200 samples)\cite{ph2}, \skincancer (206 samples)\cite{skincancer}, \ISICsix (379 samples)\cite{ISIC2016}, \ISICseven (750 samples)\cite{ISIC2017}, and \ISICeight (1,100 samples)~\cite{ISIC2018_1, ISIC2018_2}. Segmentation performance is measured using the DICE score and Jaccard Index (JAC). All foundation models are adapted for lesion segmentation by attaching a trainable segmentation head. Detailed experimental settings and implementation details are provided in the Method section. 
Overall, \derm outperform all baseline foundation models on all segmentation testing dataset in terms of both JAC and DICE ($p<0.001$), using both a simple linear segmentation head (\textbf{Fig~\ref{fig:caption_segmentation_comparison}b}) and more complex UperNet\cite{upernet} head (\textbf{Extended Data Fig.~\ref{fig:seg_result_compare_foundation}} ).    
  In the linear head setting, Compared to the strongest competing model, \panderm~\cite{panderm}, \derm achieves improvements of 2.54\% in JAC and 1.0\% in DICE score. 
\panderm outperforms \medimage when using a simple linear segmentation head, suggesting that the patch-level loss in \panderm helps the model better leverage low-level visual features extracted by the backbone, enabling strong performance even with minimal head design. In contrast, \medimage performs better with the more sophisticated UperNet head, likely because its vision-language contrastive loss enhances global semantic understanding, which is crucial for complex segmentation architectures that integrate multi-scale context.

Fig.~\ref{fig:caption_segmentation_comparison}d presents a visualization of each model’s performance across retrieval, linear classification, and segmentation tasks using heatmap. This approach allows a clear comparison of how different models balance high-level clinical tasks such as retrieval and classification with low-level visual tasks such as segmentation. \derm demonstrates strong and consistent performance across all these tasks, highlighting its robust generalizability and balanced capabilities. \lvm performs well on low-level tasks but achieves only moderate results on high-level tasks. DINOv2 shows competitive results in both classification and segmentation, but its retrieval performance is limited, likely due to the absence of domain-specific pretraining. By comparison, models such as Panderm and MedImageInsight, which benefit from relevant medical domain training, exhibit superior retrieval performance. These findings underscore the importance of both task diversity and domain adaptation in developing foundation models for dermatology, while further highlighting the advantage of \derm’s hybrid training strategy in delivering comprehensive and clinically relevant performance.





\subsection{Bias and Fairness Analysis}
Ensuring the fairness and robustness of dermatology foundation models is crucial for clinical adoption. We systematically evaluated DermINO’s diagnostic performance across key demographic subgroups, including Fitzpatrick skin types using the \onesevenkthree dataset and gender using the \dermsevenpt dataset. As shown in Fig.~\ref{fig:reader_study}a, \derm demonstrates stable and consistent performance across all Fitzpatrick types (I–VI) and both genders, indicating that the model’s accuracy does not exhibit significant demographic bias.

Importantly, even though some competing models such as PanDerm and MedImageInsight are not trained on data from Asian or darker-skinned populations, they also show relatively strong performance in Fitzpatrick types III–IV. This suggests that both the diversity of pretraining data and the design of hybrid pretraining strategies can jointly mitigate traditional sources of bias in dermatology AI. Nevertheless, further analysis across a broader range of demographic factors, including age and rare disease subtypes, will be necessary to fully ensure model equity in real-world applications.

\subsection{Federated Learning}
Given the sensitive nature of medical images, federated learning (FL) offers a privacy-preserving solution for collaborative model development across institutions. We assessed \derm’s performance under federated learning scenarios on malignant tumor diagnosis tasks using the Fitzpatrick-3, MED-Node, and \phtwo datasets. All foundation models were equipped with trainable classification heads and evaluated using AUROC and F1 scores.

As illustrated in Fig.~\ref{fig:reader_study}b, \derm consistently outperformed other models, maintaining high diagnostic accuracy and stability in federated settings. This robust performance across distributed data sources highlights \derm’s practical potential for real-world deployment, enabling secure and effective multi-center collaboration in dermatological AI without direct data sharing.

\subsection{Reader Study Evaluates \derm and Its Clinical Utility}

To evaluate the diagnostic performance of \derm and its potential as a clinical assistive tool, a reader study was conducted involving 23 dermatology image specialists, with an average of 3 years of clinical experience.

A total of 119 dermoscopic images were randomly sampled from the test set of two datasets: 80 from \classtwo and 39 from \classfive. For each image, participants select one single diagnosis they deemed most appropriate from the set of choices provided by the dataset within 20 seconds. The same set of images was also evaluated by the \derm model. Two weeks later, a second round of the reader study was conducted, in which AI predictions from \derm were provided as reference during participant diagnosis.

\begin{figure*}[hp!]
    \centering
    \includegraphics[width=1\linewidth]{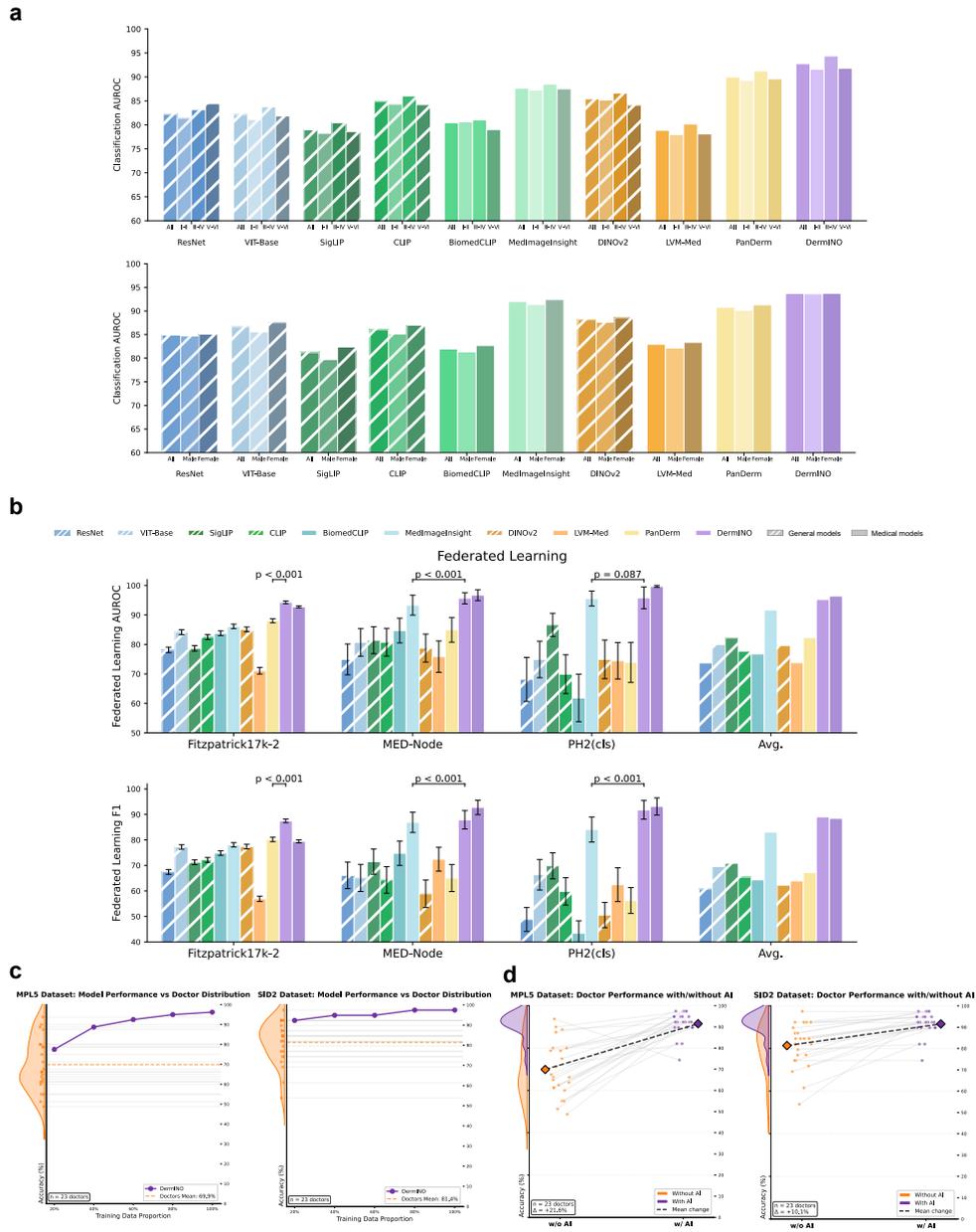}

\caption{\textbf{Evaluation of Model Fairness, Federated Learning, and Human-AI Collaboration.}
\textbf{a}, AUROC of models across FitzPatrick-3 skin types and \dermsevenpt gender subgroups shows population fairness.
\textbf{b}, Federated learning: AUROC and F1 for \derm and nine models on FitzPatrick-3, MED-Node, and PH2 show generalizability in privacy preserving scenarios.
\textbf{c}, DermINO accuracy (purple) on MPL5 and SID2 with different training sizes, compared to 23 dermatologists (violin plots, without AI).
\textbf{d}, Clinician accuracy improves with DermINO support (paired points: orange, without AI; purple, with AI).}
    \label{fig:reader_study}
\end{figure*}

As illustrated in Fig~\ref{fig:reader_study}c, in the initial round, based on the combined results from two datasets, the average diagnostic accuracy among the 23 dermatology image specialists was 73.66\%. In comparison, \derm achieved an overall diagnostic accuracy of 95.79\%, outperforming all participating readers. Specifically, on the MPL5 dataset, dermatologists achieved an average diagnostic accuracy of 69.89\%, whereas \derm attains 96.25\%, exceeding the performance of all 23 readers. On the SID2 dataset, the average reader accuracy was 81.38\%, while \derm achieved 94.87\%, outperforming 22 of the 23 participants. Furthermore, we examine how the size of the training dataset affects \derm’s performance compared to that of clinical physicians. While \derm's accuracy gradually improves with more training data, it already surpasses the diagnostic performance of experienced clinicians using only 20\% of the task-specific training set, approximately 100 dermatology images in Fig~\ref{fig:reader_study}c.

In the second round, where AI assistance was provided, almost all participants exhibited improved diagnostic performance, as demonstrated in Fig~\ref{fig:reader_study}d. Overall, the average diagnostic accuracy across both datasets increased by 17.21\%. More specifically, accuracy on the MPL5 dataset improved by 20.65\%, and on the SID2 dataset by 10.14\%. The distribution of accuracy scores before and after introducing AI assistance clearly demonstrates significant improvements.

    \section{Discussion}
In recent years, AI has achieved remarkable progress across various domains, particularly in computer vision and natural language processing. However, its application in dermatology remains relatively limited. Most existing research focuses on narrow, single-task models, such as skin cancer classification, which sometimes achieves expert-level performance but lacks the generality required to meet the diverse and complex needs of clinical dermatology. Foundation models, as a new paradigm in AI, offer the potential to overcome these limitations by leveraging large-scale unlabeled data for pretraining and enabling transferability across multiple downstream tasks. Given the high cost of data annotation and the complexity of dermatological disorders, building a task-generalizable foundation model for dermatology is both important and challenging. This challenge is compounded by the scarcity of labeled data, the wide variability of clinical tasks, and the intricacies of real-world clinical workflows.

To address these challenges, we propose a novel hybrid-supervised pretraining model, \derm, specifically designed as a foundation model for dermatology. \derm is pretrained on a total of 432,776 dermatological images, which span a broad range of sources, including public source, web source, and proprietary source. The training strategy of \derm integrates two core components. First, it adopts a hybrid-supervised learning framework, leveraging the DINO architecture to extract structural representations from unlabeled dermatology images, thereby learning generalizable features. Second, it incorporates partially supervised learning to enhance its understanding of clinically meaningful pathology using a limited number of annotated examples. Additionally, we introduce a domain knowledge-guided prototype initialization mechanism, which embeds semantic priors from dermatological disease taxonomies into the model structure. This design guides the model to learn feature representations that are more consistent with clinical semantics. Finally, we conduct a reader study to evaluate DermINO’s clinical utility.

We conducted a comprehensive evaluation of \derm on 20 dermatology datasets derived from diverse sources. The evaluation covered five representative clinical tasks from high-level to low-level: malignancy assessment, severity grading, multi-category disease diagnosis, dermatological image caption, and skin lesion segmentation. Across all tasks, \derm consistently outperformed existing state-of-the-art methods, demonstrating strong generalization capabilities across both task types and datasets. These results confirm \derm’s potential as a robust and versatile foundation model for dermatology.

\derm demonstrated strong diagnostic performance in both multi-class classification of skin tumors (\classtwo) and challenging binary classification tasks that are often difficult for clinicians to differentiate (\classfive). Notably, its diagnostic accuracy surpassed that of most participating dermatology image specialists. Furthermore, when used as an assistive tool, \derm significantly improved the diagnostic accuracy of participating clinicians across all experience levels. We also observed that dermatology image specialists demonstrated lower accuracy on the five-class \classfive dataset compared to the binary \classtwo dataset, indicating the greater complexity of multi-class classification tasks. In contrast, \derm maintained high diagnostic accuracy across both datasets. This suggests that while multi-class diagnostic tasks pose greater challenges for human clinicians, \derm is well-equipped to handle such complexity. Its consistent performance in both binary and multi-class settings underscores the robustness and adaptability of the model, highlighting its potential utility in real-world clinical scenarios where a wide spectrum of differential diagnoses must be considered. This may be attributed to \derm’s ability to capture subtle inter-class distinctions through its advanced visual feature representation. Despite these strengths, certain limitations must be acknowledged.  A subset of the evaluated images had a high magnification level, which may have impaired diagnostic clarity. For example, magnified dermoscopic images of squamous cell carcinoma and actinic keratosis can both exhibit features such as dilated follicular openings and follicular keratin plugs, making them difficult to distinguish when only a single dermoscopic image is available (Extended Data Fig.~\ref{fig:dermoscopic_scc}, \ref{fig:dermoscopic_b}). These factors likely contributed to diagnostic errors among some doctors. Interestingly, \derm was able to correctly classify such cases, suggesting a superior capacity for fine-grained visual feature recognition. This highlights the model’s potential in supporting diagnostic decisions in scenarios where human interpretation is challenged by subtle morphological similarities. 

It is also noteworthy that while most participants demonstrated significant improvements in diagnostic accuracy during the human-AI collaboration phase of the reader study, the results revealed an heightened alignment between participants’ diagnoses and the model’s predictions. When the model made mistakes, participants were also more likely to make errors on those cases, as they tend to follow the model’s suggestions. This underscores the need for careful consideration of the collaboration mode and interaction design between AI-assisted tools and human clinicians, as these factors can exert meaningful influence on diagnostic outcomes.

Despite \derm’s strong performance across multiple tasks, several limitations remain. First, the current evaluation does not cover the full spectrum of dermatological conditions, particularly rare genetic disorders and complex systemic diseases, due to limitations in available data. Second, the fairness of the model across different demographic groups, such as skin tones, genders, and age groups, has not yet been systematically assessed, which is critical for safe and equitable deployment in real-world clinical settings. Lastly, the model’s efficacy in assisting clinicians within real-world workflows has yet to be validated in practice. The true impact of \derm in clinical decision-making still requires in-depth reader studies and prospective trials.

Future work will focus on addressing these limitations by building more comprehensive and representative datasets, improving performance on long-tail diseases, and conducting multi-center clinical validation studies. In addition, we aim to establish a robust fairness evaluation framework that measures model behavior across diverse populations, thereby supporting its safe, effective, and equitable adoption in global clinical environments.

In summary, we present \derm, a hybrid-supervised dermatology foundation model that combines self-supervised and supervised learning to address the core challenges of clinical dermatology AI. We curated a large, high-quality pretraining dataset and validated the model’s capabilities across a spectrum of high-level and low-level dermatological tasks. \derm offers a new framework for foundation model development in dermatology and serves as a promising reference for future efforts in other medical specialties.

    \section{Methods}

\subsection{Implementation Details on \derm Pretraining}

\subsubsection{Overall hybrid pretraining framework}
\derm is trained using a hybrid pretraining paradigm that combines large-scale self-supervised learning with supervised training on partially labeled data. At the core of this approach is \dino~\cite{oquab2023dinov2}, a state-of-the-art self-supervised method that employs a teacher–student architecture, where a teacher network is maintained using an exponential moving average (EMA) of the student network's weights.
\derm learns robust visual representations for dermatological images using a multi-view input strategy designed to capture both global context and fine-grained local details. Specifically, for each input image, the teacher network processes $N_g$ global crops from it to encode holistic semantic information, while the student network processes both $N_g$ randomly masked global crops and $N_l$ low-resolution local crops (Fig.~\ref{fig:fig1_overview}b). 
To effectively learn from these heterogeneous views, \derm is optimized using a combination of loss functions: an image-level contrastive loss for learning high-level semantic features, a patch-level masked image modeling (MIM) loss for enhancing fine-grained low-level representations, a KoLeo regularization loss and a domain knowledge guided supervision loss.  
To further improve generalization across diverse dermatological tasks, \derm incorporates a domain-informed prototype initialization strategy, which encodes expert knowledge using the medical language model CODER~\cite{CODER} for the supervised learning objective.

We compare \derm against existing state-of-the-art foundation model pretraining paradigms including MAE~\cite{mae}, SwAV~\cite{SWAV}, MoCo v3~\cite{mocov3}, and DINOv2~\cite{oquab2023dinov2}, which represent a diverse set of pretraining objectives (Tab.\ref{tab:result_compare_methods}).  \derm consistently outperforms all baseline methods across  four categories of dermatology tasks: image retrieval, classification, segmentation and captioning. 
In addition, we conduct an ablation study to evaluate the contribution of each of the four loss components used in \derm, analyzing their impact on model performances across the same four task categories (Tab.\ref{tab:loss_ablation}).

\subsubsection{Image-level self-supervised objective}

To capture semantically meaningful representations across different granularities, DINO employs an image-level contrastive loss comprising two components. The global alignment loss aligns the student’s image-level representations of masked global crops \( \mathbf{X}^{\text{cls}}_{s,g} \in \mathbb{R}^{N_g\times D}\)  with the teacher’s outputs from unmasked global crops  \( \mathbf{X}^{\text{cls}}_{t,g} \in \mathbb{R}^{N_g\times D} \), where $D$ is the dimension of the token representations. The local-global alignment loss encourages consistency between the student’s local crop representations \( \mathbf{X}^{\text{cls}}_{s,l} \in \mathbb{R}^{N_l\times D}\) and the teacher’s global crop representations \( \mathbf{X}^{\text{cls}}_{t,g}\).

Specifically, the class tokens produced by the student network and the teacher network are first passed through their respective DINO projection heads to generate intermediate feature representations, which are then mapped into a $K$-dimensional prototype score vector. 
The student prototype scores of the input local and global crops \( p_{s,l}\in \mathbb{R}^{N_l\times K} \) and \( p_{s,g} \in \mathbb{R}^{N_g\times K} \) are computed by applying a temperature-scaled log-softmax to the projection of \( \mathbf{X}^{\text{cls}}_s \). For  \( p_{t,g} \) from the teacher network,  an additional Sinkhorn-Knopp normalization~\cite{Sinkhorn-Knopp} is adopted after the \textit{softmax} operation. Consequently, the image-level global alignment and local-global alignment loss are defined as follows:

\begin{equation}
\mathcal{L}_{\text{Image}}^{\text{Global}} = - \frac{1}{N_g(1+N_l)} \sum_{i=1}^{N_g} \sum_{k=1}^{K} p_{t,g}^{(i,k)} \log p_{s,g}^{(i,k)},
\end{equation}

\begin{equation}
\mathcal{L}_{\text{Image}}^{\text{Local-Global}} = - \frac{1}{N_g(1+N_l)} \sum_{m=1}^{N_g} \sum_{n=1}^{N_l} \sum_{k=1}^{K} p_{t,g}^{(m,k)} \log p_{s,l}^{(n,k)},
\end{equation}

\begin{equation}
\mathcal{L}_{\text{Image}} = \mathcal{L}_{\text{Image}}^{\text{Global}} + \mathcal{L}_{\text{Image}}^{\text{Local-Global}}.
\end{equation}

\derm applies the KoLeo regularization~\cite{sablayrolles2018spreading} to the image-level representations of global crops, encouraging a more uniform and well-dispersed distribution in the representation space.

\subsubsection{Patch-level self-supervised objective}
At the patch level, \derm incorporates a masked image modeling objective~\cite{patch_level_loss}, which encourages the student network to reconstruct the masked patch tokens of global crops \( \mathbf{X}^{\text{patch}}_{s,g} \in \mathbb{R}^{N_g \times N_p \times D} \) to match the corresponding output tokens produced by the teacher network  \( \mathbf{X}^{\text{patch}}_{t,g}  \in \mathbb{R}^{N_g \times N_p \times D }\), where $N_p$ denotes number of patch token in each image crop.  To obtain the patch-level prototype scores $\hat{p}_s \in \mathbb{R}^{N_g \times N_p \times K }$ and $\hat{p}_t \in \mathbb{R}^{N_g \times N_p \times K }$ from the student and teacher networks, respectively, an iBOT head~\cite{patch_level_loss} is first applied, followed by the same softmax and centering operations described in the previous section. 
\begin{equation}
\mathcal{L}_{\text{Patch}} = - \frac{1}{N_g} \sum_{i=1}^{N_g}\frac{1}{\sum_{j=1}^{N_p} {m}_{i,j}}\sum_{j=1}^{N_p} \sum_{k=1}^{K} \hat{p}_t^{(i,j,k)} \log \hat{p}_s^{(i,j,k)}, 
\end{equation}
where ${m}_{i,j}\in\{0, 1\}$ is a binary mask that select only the mask patch tokens, and \(1/(\sum_{j = 1}^{N_p}m_{i,j}) \)  is a normalization factor that re-weights the contribution of each cropped image based on the number of masked patches it contains. 

\subsubsection{Supervised objective with domain knowledge-guided prototype initialization}

As illustrated in Fig.~\ref{fig:fig1_overview}b, \derm introduces a supervision loss that aligns the image-level representations with a set of learnable prototypes that encode the knowledge and semantics of various skin diseases. 
Specifically, given the image-level representations of global image crops \( \mathbf{X}^{\text{cls}}_{s,g} \) and a set of learnable prototypes \( W \in \mathbb{R}^{N_c \times D} \) with $N_c$ as the number of prototypes, we output another set of prototype scores $p'_{s}\in\mathbb{R}^{N_g \times N_c}$ that quantifies the similarity between each image representation and the disease prototypes: 
\begin{equation}
p'_{s} =  \mathrm{Sigmoid}\left(\mathbf{X}^{\text{cls}}_{s,g} W^T\right). 
\end{equation}
Then, the supervision objective is defined as a binary cross entropy loss: 
\begin{equation}
\mathcal{L}_{sup} = -\frac{1}{N_g} \sum_{i=1}^{N_g} 
\left( \frac{1}{\sum_{j=1}^{N_c} n_{ij}} 
\sum_{j=1}^{N_c} n_{ij} \cdot (y_{ij} \log p'_{ij} + (1 - y_{ij}) \log(1 - p'_{ij}))\right), 
\label{eq:masked_bce_loss}
\end{equation}
where \( y_{ij} \in \{0,1\} \) indicates whether the $i$-th image crop is relevant to the $j$-th disease prototype. The binary mask \( n_{ij} \in \{0,1\} \) serves as an ignore indicator, set to $1$ only when the relevance label $y_{i, j}$ is available.

Each prototype vector in $W$  corresponds to a specific skin disease label, curated based on disease annotations collected from a diverse set of public dermatology datasets covering a wide spectrum of dermatological conditions. 
As shown in Fig.~\ref{fig:fig1_overview} b, we first collected all the  disease labels from multiple public sources datasets, and apply character-level normalization to unify naming conventions across datasets. 
We then perform a semantic label merging process to merge disease labels that refer to the same underlying condition. Specifically, labels with similar textual embeddings obtained using \coder~\cite{CODER} are merged, resulting in a more coherent and high-quality label set that better aligns with the underlying disease taxonomy.

To incorporate domain-specific knowledge captured by the medical language model \coder into the supervised training, we propose a domain-aware prototype initialization strategy. 
Specifically, we input the list of \( N_c \)  merged disease labels into \coder to obtain their corresponding semantic embeddings, where \( e_i \in \mathbb{R}^D \) represents the embedding of the \( i \)-th label.  These embeddings are then stacked to form an initialization matrix for the disease prototypes:

\begin{equation}
W_{\text{init}} = 
\begin{bmatrix}
e_1^\top \\
e_2^\top \\
\vdots \\
e_n^\top
\end{bmatrix}
\in \mathbb{R}^{N_c \times D}.
\end{equation}

Initializing the disease prototype matrix $W$ with \( W_{\text{init}} \) creates semantically meaningful anchors for the supervised pre-training, enabling the model to incorporate rich inter-disease semantic relationships and structure from the very beginning. 

\subsubsection{Detailed pretraining configuration}

The pre-training is performed on a high-performance computing cluster equipped with 8 NVIDIA H100 GPUs.
Using the DINO pre-trained ViT-Base model as initialization, we continue training for 100 epochs and use the model from the final epoch as our final checkpoint. The architecture consists of two ViT-Base visual encoders serving as the teacher and student networks, each containing 86 million parameters. The base learning rate is set to 2e-3, with a per-GPU batch size of 256, yielding a total effective batch size of 2048. We follow the default data augmentation settings used in \dino~\cite{oquab2023dinov2}, and adopt the AdamW optimizer. The model is trained on input images with a fixed resolution of 224×224. Unlike the original two-stage resolution scheduling strategy proposed in \dino, which gradually transitions from low-resolution to high-resolution inputs, we adopt a simplified single-stage training scheme with fixed resolution. This design choice reduces computational overhead while maintaining competitive performance.

\subsection{Implementation Details on Downstream Evaluation}
For all downstream tasks, we employed a non\-parametric bootstrap with 1,000 replicates. From these replicates we calculated the bootstrap mean \(\mu\) of each performance metric and its bootstrap variance \(v\). A \(95\%\) Confidence Interval(CI) was then constructed with the lower and upper bounds defined as \(\mu - v\) and \(\mu + v\), respectively. To assess statistical significance between models, the \(1{,}000\) bootstrap estimates were treated as paired observations, and two-sided paired \(t\)-tests were performed to obtain the corresponding \(p\)-values.

\subsubsection{Classification for dermatology clinical tasks}

For the image retrieval task, we employed a zero-shot evaluation protocol, designed to assess the generalization capability of pre-trained visual encoders without any task-specific fine-tuning. In this setup, the backbone of each model was frozen and used solely for visual feature extraction, ensuring that the evaluation reflects the inherent quality of the pre-trained representations. The extracted features from each image were projected into a shared embedding space, where semantic similarity could be measured. Retrieval was performed using a $K$-nearest neighbor ($K=5$) search, identifying the top 5 closest training samples to each test image. To compute similarity between image embeddings, we used cosine similarity, which is a standard metric in high-dimensional embedding spaces for assessing semantic alignment. This allowed us to effectively rank the neighbors based on their closeness to the query image. This design enabled us to quantitatively evaluate each model’s ability to retrieve semantically relevant images from a diverse set of diagnostic categories, relying purely on their pre-trained visual knowledge, without introducing any supervision or adaptation bias.

In the image classification task, we extended each model by attaching a trainable classification head, typically a fully connected (linear) layer, on top of the frozen pre-trained visual encoder. 
The classification head was trained using the cross-entropy loss function. The optimization was performed using the Adam optimizer. We initialize training with a learning rate of 5e-3 and adopt a cosine annealing schedule~\cite{CosineAnnealing} for learning rate decay. Each model was trained for 50 epochs, and all training was conducted under uniform settings across models, ensuring a fair comparison of performance and isolating the effect of pre-trained feature quality rather than hyperparameter variance.

Model performance for both retrieval and classification tasks was quantitatively assessed using macro-averaged AUROC \cite{AUROC} and macro-averaged F1 scores\cite{F1}.

\subsubsection{Image caption for dermatology clinical tasks}
For the image captioning task, visual features were extracted using a frozen image encoder backbone, and were subsequently passed through a trainable linear projection layer, which mapped them into the textual embedding space compatible with the language model. 
This projection layer was optimized with an initial learning rate of \(4 \times 10^{-4}\) to facilitate effective alignment with the downstream captioning objective. The resulting projected embeddings were then fed into a frozen medical language model, \bio~\cite{biomistral}, which was adapted to the captioning task using low-rank adaptation (LoRA)~\cite{Lora}. The LoRA parameters were trained with a smaller learning rate of \(8 \times 10^{-5}\) to ensure fine-grained adaptation while preserving the pretrained knowledge of the language model.
The model was trained using the AdamW optimizer for 2 epochs. We adopted a comprehensive set of standard evaluation metrics to assess the quality of generated captions, including BLEU-1, BLEU-2, BLEU-4~\cite{Bleu}, METEOR~\cite{meteor}, CIDEr~\cite{cider}, ROUGE-1, ROUGE-2, and ROUGE-L~\cite{rouge}.

\subsubsection{Segmentation for dermatology tasks}

We evaluated the model's performance on the dermatology segmentation task by adopting a frozen visual encoder backbone coupled with two types of lightweight, trainable segmentation heads \cite{upernet}. For the first segmentation head, synchronized batch normalization is first applied to the input features, followed by spatial dropout with a probability of 0.1 to enhance generalization. A final $1 \times 1$ convolutional layer then maps the features to the desired number of output classes, producing dense segmentation logits. 
The second segmentation head adopts a more structured decoder based on the UperNet architecture~\cite{upernet}. It aggregates multi-scale features from different stages of the backbone using a pyramid pooling module, followed by feature fusion and refinement. This design enables the model to capture both local details and global contextual information, which is especially beneficial for complex dermatological patterns. The UPerNet head concludes with a $1 \times 1$ convolutional layer to generate the final segmentation logits.

During training, all input images and ground-truth masks were uniformly resized to $224 \times 224$ to ensure consistent spatial alignment and computational efficiency. Only the segmentation head was trained, while the backbone remained frozen throughout the process, allowing us to isolate and evaluate the quality of the pretrained visual features. The model was optimized using the Adam optimizer, with an initial learning rate of $1 \times 10^{-3}$ applied to the segmentation head. The learning rate was decayed following a cosine annealing schedule over the course of 20 training epochs, promoting smooth convergence without the need for manual tuning.

To assess segmentation quality, we employed two widely used overlap-based evaluation metrics: the Jaccard Index (JAC)~\cite{jac} and the Dice Similarity Coefficient (DICE)~\cite{dice}, both of which are standard for evaluating pixel-wise agreement in semantic segmentation tasks. These metrics capture complementary aspects of prediction quality: Jaccard focuses on the ratio between intersection and union, while Dice emphasizes the harmonic balance between precision and recall at the pixel level.


\subsubsection{Federated learning for dermatology tasks}
For the federated learning task, we conducted experiments using the Flower~\cite{flwr} framework on three datasets, \onesevenktwo, \mednode and \phtwo. We follow the same data partitioning strategy as described in the classification section. In our experiments, a frozen image encoder backbone was used to extract visual features, which were then fed into a trainable linear projection layer. Each model was locally trained on its respective dataset for 2 epochs with an initial learning rate of $2 \times 10^{-4}$, using the Adam optimizer and a cosine annealing learning rate scheduler. After local training, model parameters were aggregated, and this federated learning process was repeated for 2 global epochs. We report the performance of the aggregated model across all three datasets. The evaluation metrics used are AUROC and F1 scores.

\subsection{Pretraining Datasets}

We curated an extensive pretraining dataset comprises 432,776 multi-source skin images to develop DermINO. This diverse pretraining dataset is obtained from three sources: 232,215 (53.7\%) high-resolution dermatology images from proprietary source; 95,999 (22.2\%) images obtained from the public LESION 130K dataset~\cite{leision130k}, originally crawling from webset; 104,562 (24.2\%) images from the public clinic image dataset with annotated labels for supervised model pretraining. This multi-source dataset affiliated with the hybrid pretraining strategy provides a comprehensive representation of skin lesions, enabling the model to learn robust features across different downstream tasks. The ablation study in Tab.~\ref{tab:pretrain_downstream_datasets} evaluates the impact of pretraining dataset composition by using three progressively enriched datasets with increasing diversity of data sources. The results demonstrate how different dataset sources influence the performance of the pretrained foundation model across four categories of dermatology tasks.

\subsubsection{Proprietary source}

\textbf{\hospital} The proprietary dataset, consisting of 232,215 high-resolution clinical and dermoscopic images without diagnostic annotations from 44,668 individuals, was obtained from the Beijing China-Japan Friendship Hospital. These images, accumulated over a decade by the hospital’s dermatology department, capture a broad spectrum of real-world clinical cases and provide rich visual diversity to support model pretraining. The overall mean age was 41.88 years (SD = 16.73 years). Males accounted for 44.82\% of the population, with a mean age of 42.01 years (SD = 17.05 years), while females accounted for 55.18\%, with a mean age of 41.78 years (SD = 16.48 years).

\subsubsection{Public source}

\textbf{ISIC-Duplicate.~\cite{isic_Duplicate}} The International Skin Imaging Collaboration (ISIC) datasets comprise tens of thousands of dermoscopic images, each annotated with gold-standard lesion diagnosis metadata. Following the strategy proposed, we identify and remove duplicate images and adopt the curated version of the dataset recommended for researchers working with ISIC data, which contains 57,062 images.

 \textbf{SCIN.~\cite{SCIN}} The SCIN dataset was collected through a voluntary, consent-based image donation app from Google Search users in the United States. It comprises contributions from over 5,000 volunteers, totaling more than 10,000 images of common skin conditions. From this dataset, we curated 10,379 images for use in our pretraining.

\textbf{SD-198.~\cite{sd-198}} The SD-198 benchmark dataset contains 6,584 images across 198 distinct skin conditions, including various types of eczema, acne, and cancerous lesions. These images exhibit significant variation in size, color, shape, and structure. We used the entire dataset directly for our pretraining.

\textbf{\pad.~\cite{PAD_data}} The PAD-UFES-20 dataset contains 2,298 clinical images from 1,641 skin lesions across 1,373 patients, covering six types of skin conditions (three skin cancers and three benign lesions). Each sample includes a clinical image and up to 26 metadata features, such as patient age, lesion location, and skin type. From this dataset, we curated 1,149 images for use in our pretraining.

\textbf{DermNet.~\cite{Dermnet}} The DermNet dataset contains over 25,000 de-identified clinical images spanning 600+ skin conditions, reviewed by medical experts. It provides a diverse and reliable resource to support AI research in dermatology. From this dataset, we curated 18,856 images for use in our pretraining. We have obtained explicit permission from the dataset authors for research use in our pretraining and model release.

\textbf{Downstream train sets.} To further enhance the generalization ability of the pretraining model on downstream tasks, we incorporated the training sets of several downstream datasets into our pretraining data. These include \acnezerofour~\cite{acne04}, \ddi~\cite{ddi}, \dermsevenpt~\cite{derm7pt}, \phtwo~\cite{ph2}, \onesevenkthree~\cite{fit}, and \mednode~\cite{med_node}, contributing a total of 10,532 images to the pretraining stage.

\subsubsection{Web source}

\textbf{\web} Web source datasets consisted of 95,999 images obtained from the LESION 130k dataset, which were collected using the URLs provided by LESION 130ks~\cite{leision130k}. These images were originally acquired through large-scale web crawling with dermatology-related keywords, spanning 18,482 websites across approximately 80 countries. The curated subset was selected to maximize visual diversity; however, no diagnostic labels were associated with these images.

In total, our pretraining dataset consists of 432,776 dermatological images, encompassing both labeled and unlabeled samples collected from proprietary source, public source, and web source. This diverse and large-scale corpus provides a rich foundation for robust representation learning, enhancing the model’s ability to generalize across a wide range of dermatological tasks and disease presentations.

\subsection{Evaluation Datasets}

\subsubsection{Classification for dermatology clinical datasets}

\textbf{\classtwo}: This proprietary dataset, provided by China-Japan Friendship Hospital, consists of 1,022 images spanning five classes of malignant skin cancer: melanoma, Bowen disease, squamous cell carcinoma, basal cell carcinoma, and actinic keratosis. The dataset is stratified into 50\% training and 50\% test sets.

\textbf{\ddi}~\cite{ddi}: This public dataset \ddi consists of 656 images spanning two classes: benigh and malignant. The images included in the \ddi dataset were retrospectively selected from reviewing pathology reports in Stanford Clinics from 2010-2020. The dataset is stratified into 50\% training and 50\% test sets.

\textbf{\onesevenktwo}~\cite{fit}: The Fitzpatrick17k dataset consists of 16,577 clinical images sourced from two dermatology atlases—DermaAmin and Atlas Dermatologico. The \onesevenktwo dataset is derived from Fitzpatrick17k by selecting images from two specific categories: benign and malignant. It contains a total of 4,497 images. The data split for these two selected categories is consistent with the split used in the \onesevenkthree dataset.

\textbf{\mednode}~\cite{med_node}: The public \mednode dataset consists of 70 Melanoma and 100 Naevus images from the digital image archive of the Department of Dermatology of the University Medical Center Groningen (UMCG). The dataset is stratified into 50\% training and 50\% test sets. 

\textbf{\phtwo}~\cite{ph2}: The public dermoscopic \phtwo dataset consists of 80 common nevi, 80 atypical nevi, and 40 melanoma images, acquired at the Dermatology Service of Hospital Pedro Hispano in Matosinhos, Portugal. We grouped the 80 common nevi and 80 atypical nevi into a single benign class, forming in a binary classification dataset with 160 benign and 40 melanoma images. The dataset is stratified into 50\% training and 50\% test sets.

\textbf{\classfourone}: This proprietary dataset, provided by China-Japan Friendship Hospital, contains 500 images categorized into two classes: stable-phase melasma and active-phase melasma. The dataset is stratified into 50\% training and 50\% test sets.

\textbf{\classfourtwo}: This proprietary dataset, provided by China-Japan Friendship Hospital, consists of 500 images spanning two classes: pigmentary-type melasma and mixed-type melasma. The dataset is stratified into 50\% training and 50\% test sets.

\textbf{\acnezerofour}~\cite{acne04}: The public dataset \acnezerofour consists of 1,457 images spanning four classes: mild, modetate, severe, and very severe acne. The dataset is stratified into 50\% training and 50\% test sets.

\textbf{\classone}: This proprietary dataset, provided by China-Japan Friendship Hospital, contains 2,000 images categorized into six classes: solar lentigo, seborrheic keratosis, pigmented nevus, hemangioma, psoriasis and others. The dataset is stratified into 50\% training and 50\% test sets.

\textbf{\classfive}: This proprietary dataset, provided by China-Japan Friendship Hospital, contains 811 images categorized into two classes: scalp psoriasis and scalp seborrheic dermatitis. The dataset is stratified into 50\% training and 50\% test sets.

\textbf{\classeight}: This proprietary dataset, provided by China-Japan Friendship Hospital, contains 516 images categorized into four classes of viral wart: common wart, condyloma acuminatum, verruca plana and verruca plantaris. The dataset is stratified into 50\% training and 50\% test sets.

\textbf{\dermsevenpt}~\cite{derm7pt}: The public \dermsevenpt dataset contains 2,013 images categorized into five classes: 
basal cell carcinoma (BCC), melanoma (MEL), melanocytic nevus (NEV), seborrheic keratosis (SK), and miscellaneous (MISC). The dataset is stratified into 50\% training and 50\% test sets.

\textbf{\onesevenkthree}~\cite{fit}: The \onesevenkthree dataset is derived from Fitzpatrick17k by adopting the official three-class categorization provided by the dataset: benign, malignant, and non-neoplastic. It consists of 16,577 images and is stratified into 50\% training and 50\% test sets.

\textbf{\onesevenknine}~\cite{fit}: The \onesevenknine dataset is derived from Fitzpatrick17k by adopting the official nine-class categorization provided by the dataset: benign dermal, benign epidermal, benign melanocyte, genodermatoses, inflammatory, malignant cutaneous lymphoma, malignant dermal, malignant epidermal, and malignant melanoma. It consists of 16,577 images, with the same 50\% training and 50\% test split as used in the \onesevenkthree dataset, but with labels corresponding to the nine-class categorization.

\subsubsection{Caption for dermatology clinical datasets}
\textbf{\skincap}~\cite{skincap}: The public \skincap dataset consists of 4,000 images sourced from the Fitzpatrick17k~\cite{fit} and Diverse Dermatology Images datasets~\cite{ddi}. Among them, 3,600 images are used for training and 400 images are reserved for testing. Note that we ensured that the data used for pretraining did not appear in the training set of the current task.

\subsubsection{Segmentation for dermatology datasets}
\textbf{\segmentphtwo}~\cite{ph2}: The public \segmentphtwo dataset, which corresponds to the same source as the previously mentioned \phtwo dataset, includes 200 images, each accompanied by binary segmentation labels. In this setting, we adopt the same data partitioning strategy as used in the \phtwo dataset.

\textbf{\skincancer}~\cite{skincancer}: The public dataset \skincancer is maintained by VISION AND IMAGE PROCESSING LAB, University of Waterloo. The images of the dataset were extracted from the public databases DermIS and DermQuest, along with manual segmentations of the lesions. We randomly selected 50\% of the images for training, with the remaining 50\% used for testing.

\textbf{\ISICsix, \ISICseven, \ISICeight}~\cite{ISIC2016, ISIC2017, ISIC2018_1, ISIC2018_2}: These three public datasets consist of 379, 750, and 1,100 images, respectively, each paired with corresponding binary segmentation labels. These datasets are constructed by combining the validation and test sets from the lesion segmentation tasks of the ISIC Challenges in 2016, 2017, and 2018. For each dataset, we randomly split 50\% of the images for training and 50\% for testing.

\subsubsection{Federated learning for dermatology datasets}
In the federated learning section, we utilize the same three dermatology clinical datasets introduced in the previous \textit{Classification for dermatology clinical datasets} section, namely \mednode, \phtwo, and \onesevenktwo.

    \bibliographystyle{naturemag}
    \bibliography{sn-bibliography}

    \section*{Acknowledgements}
This work was supported in part by the National Key R\&D Program of China under Grant No.2023YFA1008600, in part by Beijing Chaoyang Digital Health Proof of Concept Project (NO.2025SLZD010),National High Level Hospital Clinical Research Funding (No.2023-NHLHCRF-YXHZ-ZRZD-06) and Beijing Municipal Science and Technology Commission Medicine Collaborative Science and Technology Innovation Research Project (No.Z191100007719001).

\section*{Ethics Approval}
Reviewed and approved by the Clinical Research Ethics Committee of China-Japan Friendship Hospital (2023-KY-218 and 2020-130-K83).



\section*{Extended Data}
\FloatBarrier

\begin{extendeddatatable*}[t]
    \centering
    \small
    \setlength{\tabcolsep}{2pt}
    \caption{\textbf{Ablation Study on Loss Functions.}
    Ablation study results evaluating the contributions of the original \dino loss functions and the proposed knowledge guidance loss. The evaluation is performed across three clinical tasks: malignancy assessment, severity grading, and multi-category diagnosis, using both retrieval and classification settings, with AUROC and F1 score as performance metrics. In addition, skin lesion segmentation performance is assessed using JAC and DICE scores, under two segmentation head configurations: a linear segmentation head and the UperNet head. Furthermore, dermatological image captioning is evaluated using METEOR and CIDEr metrics.
    }
    \resizebox{\textwidth}{!}{
    \begin{tabular}{@{}c c c c c c c c c  c c c c c c c @{}}
        \toprule
        \multicolumn{2}{c}{Retrieval} &
        \multicolumn{2}{c}{Classification} &
        \multicolumn{2}{c}{\makecell{Segmentation \\ Linear}} & 
        \multicolumn{2}{c}{\makecell{Segmentation \\ UperNet}} & 
        \multicolumn{2}{c}{Caption} & 
        \multicolumn{2}{c}{\makecell{Federated \\ Learning}} & 

        \multirow{2}{*}{\makecell{Image \\ Level}} &
        \multirow{2}{*}{\makecell{Patch \\ Level}} &
        \multirow{2}{*}{\makecell{Regularization \\ Level}} &
        \multirow{2}{*}{\makecell{Knowledge \\ Guidance}} 
        \\
        \cmidrule(lr){1-2}
        \cmidrule(lr){3-4} 
        \cmidrule(lr){5-6} 
        \cmidrule(lr){7-8} 
        \cmidrule(lr){9-10} 
        \cmidrule(lr){11-12} 
        AUROC & F1 & AUROC & F1 & JAC& DICE & JAC & DICE & METEOR & CIDEr & AUROC & F1 &  & & & \\
        \midrule

        91.08 & 80.35 & 93.98 & 82.57 & 86.57 & 89.22 & 91.45 & 93.6  & 0.3047 & 0.1574 & \textbf{96.19} & 87.15 & $\checkmark$ &  &  &                \\
        91.31 & \textbf{82.21} & 94.00 & 83.60 & \textbf{88.29} & \textbf{90.65} & \textbf{91.66} & \textbf{93.78} & 0.3116 & \textbf{0.1870} & \textbf{96.19} & \textbf{89.60} & $\checkmark$ & $\checkmark$ &  &          \\
        \rowcolor{lightgray}
        \textbf{91.42} & 81.77 & \textbf{94.48} & \textbf{84.26} & 88.03 & 90.40 & 91.56 & 93.70 & \textbf{0.3119} & 0.1735 & 95.25 & 89.06 & $\checkmark$ & $\checkmark$ & $\checkmark$ & $\checkmark$   \\
        \midrule
        86.83 & 71.30 & 91.91 & 78.74 & 87.26 & 89.78 & 91.42 & 93.60 & 0.2905 & 0.1477 & 80.79 & 71.60 & $\checkmark$ &  $\checkmark$      & $\checkmark$ &  \\
        \rowcolor{lightgray}
        \textbf{91.42} & \textbf{81.77} & \textbf{94.48} & \textbf{84.26} & \textbf{88.03} & \textbf{90.40} & \textbf{91.56} & \textbf{93.70} & \textbf{0.3119} & \textbf{0.1735} & \textbf{95.25} & \textbf{89.06}
         & $\checkmark$ & $\checkmark$ & $\checkmark$ & $\checkmark$ \\
        

        \bottomrule
    \end{tabular}}

    \label{tab:loss_ablation}
\end{extendeddatatable*}

\begin{extendeddatatable*}[t]
    \centering
    \small
    \caption{\textbf{Ablation Study on Pretraining Method}.
        Comparative experiments on pretraining methods, evaluating four representative approaches: \mae, \swav, \moco, and our proposed \derm. The evaluation covers five dermatology-related tasks: malignancy assessment, severity grading, multi-category diagnosis, dermatological image captioning, and skin lesion segmentation.
    }
    \label{tab:result_compare_methods}
    \resizebox{\textwidth}{!}{
    \begin{tabular}{@{}c cc cc cc cc cc cc cc@{}}
        \toprule
        \multirow{2}{*}{\makecell{Pretrain \\ method}} &
        \multicolumn{2}{c}{Retrieval} &
        \multicolumn{2}{c}{Classification} &
        \multicolumn{2}{c}{\makecell{Segmentation \\ Linear}} & 
        \multicolumn{2}{c}{\makecell{Segmentation \\ UperNet}} & 
        \multicolumn{2}{c}{Caption} & 
        \multicolumn{2}{c}{\makecell{Federated \\ Learning}} \\
        
        \cmidrule(lr){2-3} 
        \cmidrule(lr){4-5} 
        \cmidrule(lr){6-7} 
        \cmidrule(lr){8-9} 
        \cmidrule(lr){10-11} 
        \cmidrule(lr){12-13} 
        & AUROC & F1 & AUROC & F1 & JAC & DICE & JAC & DICE & METEOR & CIDEr & AUROC & F1 \\
        \midrule
        
        \mae & 82.12 & 63.82 & 86.56 & 64.63 & 86.73 & 89.36 & 90.40 & 92.60 & 0.2784 & 0.1141 & 79.85 & 66.87 \\
        \moco & 85.60 & 70.30 & 88.91 & 70.84 & 87.53 & 90.00 & 90.83 & 93.07 & 0.2900 & 0.1294 & 77.83 & 69.48 \\
        \swav & 85.03 & 71.21 & 89.73 & 73.41 & 83.72 & 86.96 & 90.00 & 92.45 & 0.2794 & 0.1015 & 80.44 & 71.19 \\
        \dino & 82.85 & 63.22 & 90.34 & 74.11 & 84.77 & 87.97 & 90.12 & 92.46 & 0.2950 & 0.1248 & 79.59 & 62.26 \\
        \rowcolor{lightgray}
        \textbf{\derm} & \textbf{91.42} & \textbf{81.77} & \textbf{94.48} & \textbf{84.26} & \textbf{88.03} & \textbf{90.40} & \textbf{91.56} & \textbf{93.70} & \textbf{0.3119} & \textbf{0.1735} & \textbf{95.25} & \textbf{89.06} \\
        
        \bottomrule
    \end{tabular}

    }
\end{extendeddatatable*}



\begin{extendeddatatable*}[t]
    \centering
    \small

    \caption{\textbf{Ablation Study on Dataset Source}.
        We conducted ablation studies on the pretraining datasets, comparing data from three distinct sources: proprietary, public, and web. The evaluation was performed across five dermatology-related tasks: malignancy assessment, severity grading, multi-category diagnosis, dermatological image captioning, and skin lesion segmentation.
    }
    \resizebox{\textwidth}{!}{
    \begin{tabular}{@{}c c c c c c c  c c c c c c c c@{}}
        \toprule
        \multicolumn{2}{c}{Retrieval} &
        \multicolumn{2}{c}{Classification} &
        \multicolumn{2}{c}{\makecell{Segmentation \\ Linear}} & 
        \multicolumn{2}{c}{\makecell{Segmentation \\ UperNet}} & 
        \multicolumn{2}{c}{Caption} & 
        \multicolumn{2}{c}{\makecell{Federated \\ Learning}} & 
        \multirow{2}{*}{Public} &
        \multirow{2}{*}{Web} &
        \multirow{2}{*}{Proprietary} 
        \\
        \cmidrule(lr){1-2} 
        \cmidrule(lr){3-4} 
        \cmidrule(lr){5-6} 
        \cmidrule(lr){7-8} 
        \cmidrule(lr){9-10} 
        \cmidrule(lr){11-12} 
        
        AUROC& F1& AUROC & F1 & JAC & DICE & JAC & DICE & METEOR  & CIDEr & AUROC & F1\\
        \midrule      
        89.41 & 77.13 & 92.31 & 79.11 & 87.57 & 90.03 & 91.61 & 93.70 & 0.2983 & 0.1684 & 93.39 & 86.05 & \checkmark \\
        90.28 & 78.27 & 92.94 & 81.21 & 87.87 & 90.31 & 91.56 & 93.71 & 0.3011 & 0.1507 & 93.68 & 86.76 & \checkmark & \checkmark \\
        \rowcolor{lightgray}
        \midrule      
        \textbf{91.42} & \textbf{81.77} & \textbf{94.48} & \textbf{84.26} & \textbf{88.03} & \textbf{90.40} & \textbf{91.56} & \textbf{93.70} & \textbf{0.3119} & \textbf{0.1735} & \textbf{95.25} & \textbf{89.06} & \checkmark & \checkmark & \checkmark \\
        \bottomrule
    \end{tabular}}
    \label{dataset ablation}
\end{extendeddatatable*}

\begin{extendeddatatable*}[ht]
\centering
\caption{\textbf{Dataset Overview for Pretraining and Downstream Tasks in DermINO.} Summary of datasets used for pretraining and downstream evaluation. The \textit{Modalities} column encompasses three types: Dermoscopic, Clinical, and Diverse, with \textit{Diverse} referring to datasets that include both Dermoscopic and Clinical images.}
\scriptsize 
\renewcommand{\arraystretch}{1.2} 
\begin{tabularx}{\textwidth}{@{}c X X c c@{}}
\toprule
\textbf{\makecell{Pretrain \\ source}} & \textbf{Datasets} & \textbf{Modalities} & \textbf{\makecell{Nums for\\ Pretraining}} & \textbf{\makecell{Labels for\\ Pretraining}} \\
\midrule
Proprietary & CSID-CJFH & Diverse & 232215 & 0  \\
\cdashline{1-5}[0.5pt/2pt]
Public & ISIC-Duplicate & Dermoscopic & 57062 & 57062  \\
Public & SCIN & Diverse & 10379 & 6505  \\
Public & SD-198 & Diverse & 6584 & 6584  \\
Public & Dermnet & Clinical & 18856 & 18856  \\
Public & PAD-UFES-20 & Clinical & 1149 & 1149  \\
Public & \makecell[l]{Downstream \\ train sets} & Diverse & 10532 & 10532  \\
\cdashline{1-5}[0.5pt/2pt]
WEB & LESION130K & Diverse & 95999 & 0  \\
\midrule
\textbf{\makecell{Downstream \\ source}} & \textbf{Datasets} & \textbf{Modalities} & \textbf{\makecell{Downstream \\Data nums}} & \textbf{\makecell{Downstream \\ tasks}} \\
\midrule
Proprietary & MPL5 & Dermoscopic & 1002 & \multirow{5}{*}{\makecell{Malignancy \\ assessment}} \\
Public & DDI & Clinical & 656 & \\
Public & Fitzpatrick17k-2 & Clinical & 4497 & \\
Public & MED-Node & Clinical & 170 & \\
Public & PH2(cls) & Dermoscopic & 200 & \\
\cdashline{1-5}[0.5pt/2pt]
Proprietary & MSD2 & Dermoscopic & 500 & \multirow{3}{*}{Severity Grading} \\
Proprietary & MTD2 & Dermoscopic & 500  & \\
Public & ACNE04 & Clinical & 1457  & \\
\cdashline{1-5}[0.5pt/2pt]
Proprietary & GLD6 & Dermoscopic & 2000  & \multirow{6}{*}{\makecell{Multi-category \\ Disease diagnosis}} \\
Proprietary & SID2 & Dermoscopic & 811  & \\
Proprietary & VWCD4 & Dermoscopic & 516  & \\
Public & Derm7pt & Dermoscopic & 2013  & \\
Public & Fitzpatrick17k-3 & Clinical & 16577  & \\
Public & Fitzpatrick17k-9 & Clinical & 16577  & \\
\cdashline{1-5}[0.5pt/2pt]
Public & SkinCAP & Clinical & 4000 &  Caption \\
\cdashline{1-5}[0.5pt/2pt]
Public & PH2(seg) & Dermoscopic & 200 &  \multirow{5}{*}{Segmentation} \\
Public & Skincancer & Diverse & 206  & \\
Public & ISIC2016 & Dermoscopic & 379  & \\
Public & ISIC2017 & Dermoscopic & 750  & \\
Public & ISIC2018 & Dermoscopic & 1100  & \\
\bottomrule
\end{tabularx}

\label{tab:pretrain_downstream_datasets}
\end{extendeddatatable*}

\begin{extendeddatatable*}[t]
\centering
\small
\setlength{\tabcolsep}{6pt}
\caption{\textbf{Skin Type and Gender Distributions Across Datasets.} Skin type distribution for \onesevenkthree dataset and gender distribution for \dermsevenpt dataset}
\resizebox{\textwidth}{!}{
\begin{tabular}{@{} c  c c c | c  c c @{}}
\hline
\onesevenkthree & \makecell{Num for skin\\ type I--II} & \makecell{Num for skin\\ type III--IV} & \makecell{Num for skin\\ type V--VI} 
& \dermsevenpt & \makecell{Num for\\ male} & \makecell{Num for\\ female} \\
\hline
Benign          & 554  & 422  & 90   & BCC  & 19  & 23  \\
Malignant       & 589  & 387  & 108  & NEV  & 281 & 295 \\
non-neoplastic  & 2739 & 2214 & 881  & MEL  & 110 & 142 \\
               &      &      &      & MISC & 43  & 55  \\
               &      &      &      & SK   & 24  & 21  \\
\hline
\end{tabular}}
\end{extendeddatatable*}

\begin{extendeddatafigure*}[p!]
    \centering
    \includegraphics[width=1\linewidth]{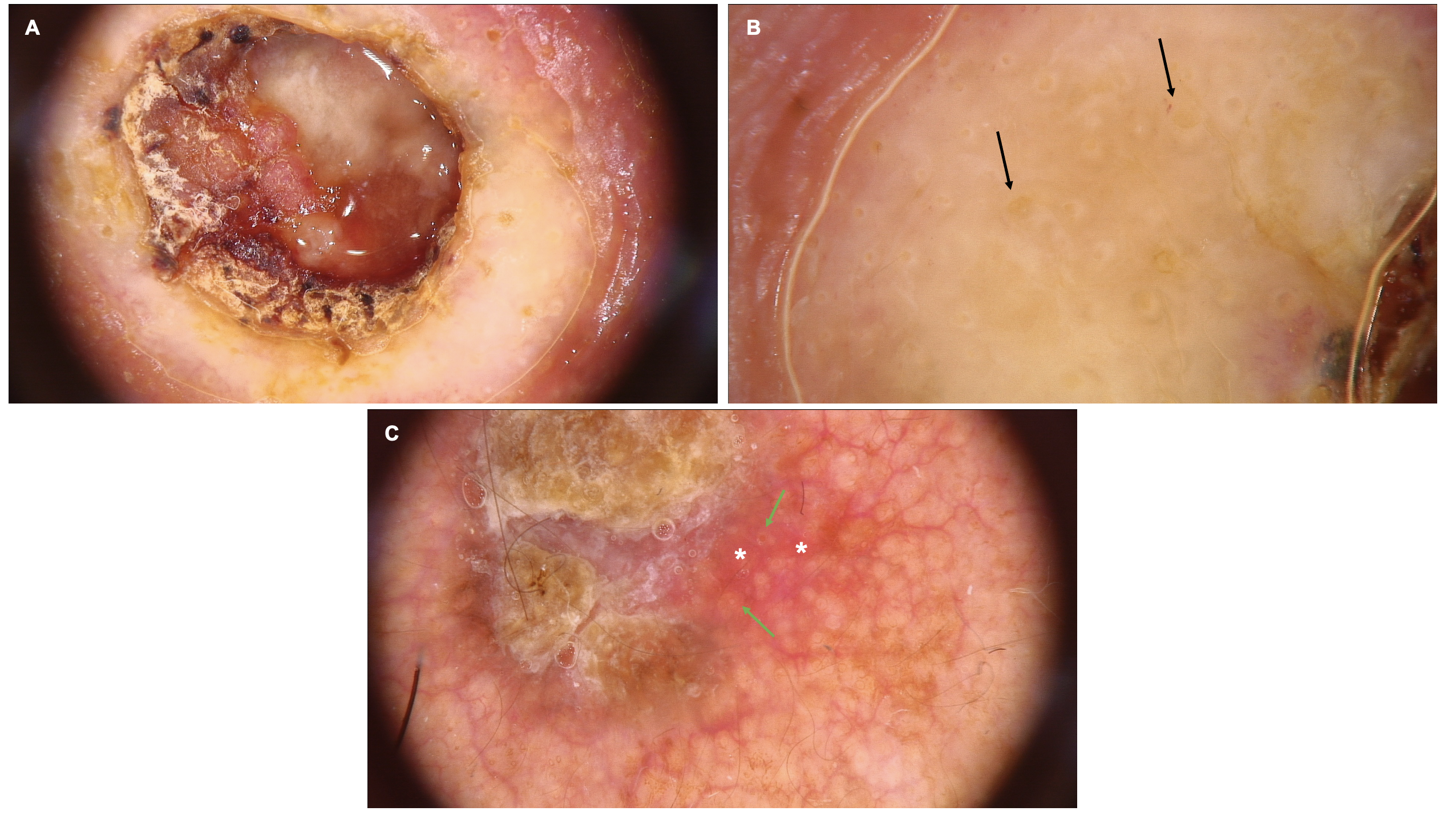}
    \caption{\textbf{High‑magnification dermoscopic similarity between squamous cell carcinoma and actinic keratosis.} (A) Low‑magnification dermoscopic image of squamous cell carcinoma (×20). (B) High‑magnification dermoscopic image of squamous cell carcinoma (×50); the black arrows indicate dilated follicular openings with follicular keratin plugs. (C) High‑magnification dermoscopic image of actinic keratosis (×30); the green arrows highlight dilated follicular openings with follicular keratin plugs, and the white asterisks indicate perifollicular linear vessels. At high magnification, both lesions display overlapping dermoscopic features, particularly dilated follicular openings and keratin plugs, which may lead to diagnostic confusion when only a single dermoscopic image is available.}
    \label{fig:dermoscopic_scc}
\end{extendeddatafigure*}

\begin{extendeddatafigure*}[htbp!]
    \centering
    \includegraphics[width=1\linewidth]{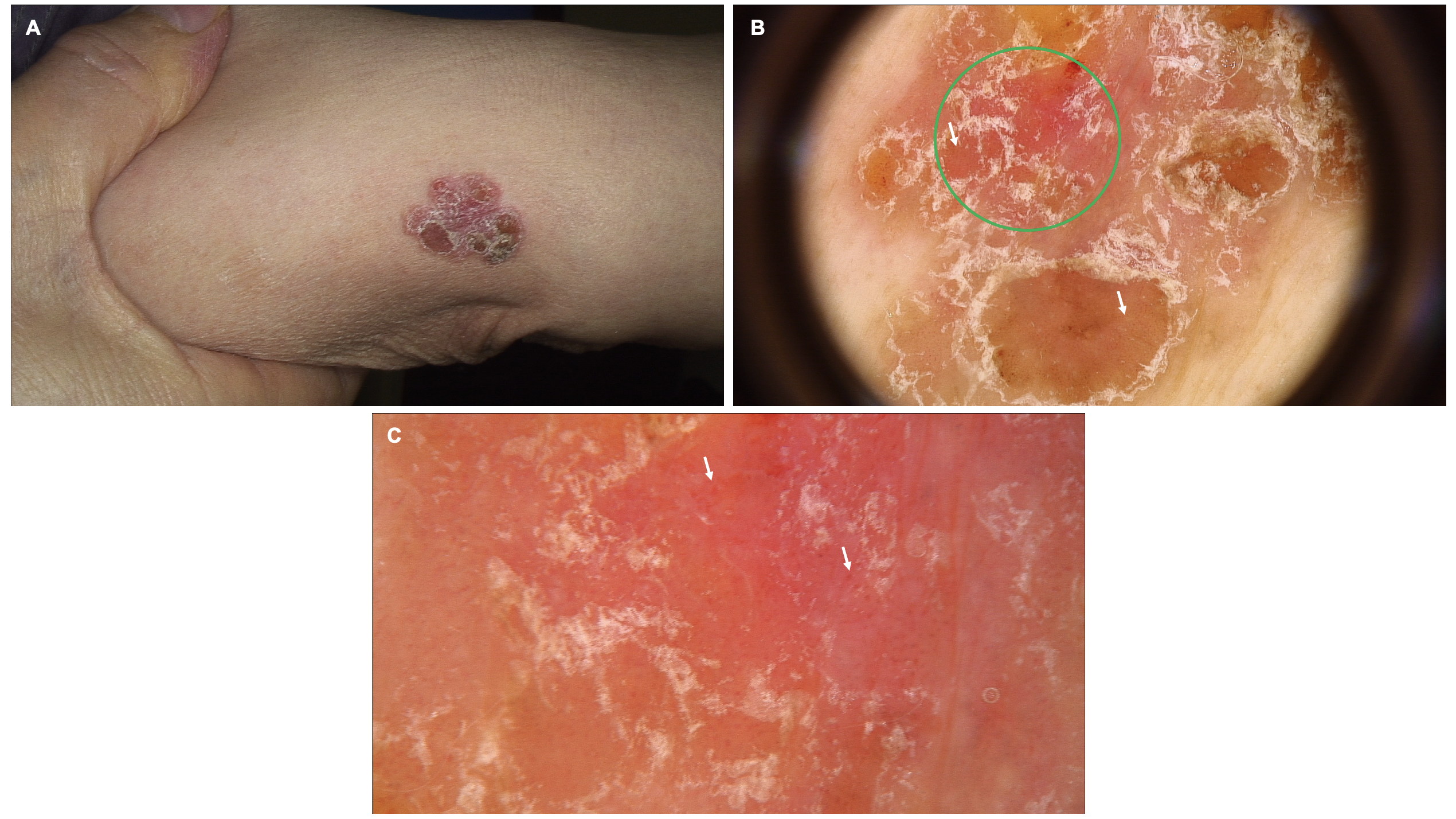}
    \caption{\textbf{Clinical and dermoscopic features of a Bowen’s disease lesion resembling psoriasis.} (A) Clinical appearance of a Bowen’s disease lesion. (B) Low-magnification dermoscopic image (×20); a reddish background with focally distributed dotted vessels (white arrows). (C) High-magnification dermoscopic image (×50); magnified view of the green-circled area in panel B, revealing a reddish background with uniformly distributed glomerular vessels (white arrows). This appearance is similar to the high-magnification dermoscopic features of psoriasis and may lead to misdiagnosis as psoriasis.}
    \label{fig:dermoscopic_b}
\end{extendeddatafigure*}

\begin{extendeddatafigure*}[htbp!]
    \centering
    \includegraphics[width=1\linewidth]{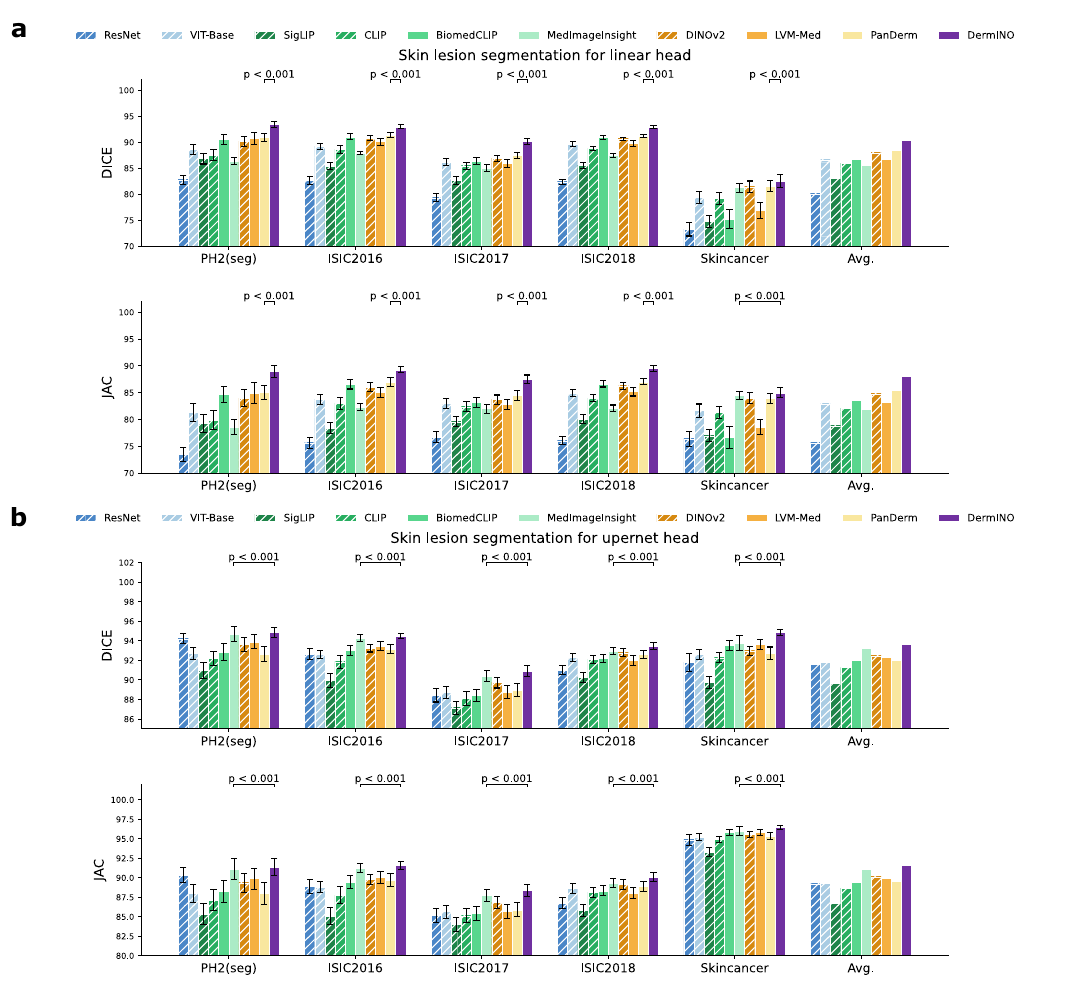}
    \caption{\textbf{DermINO Performance on Segmentation.} \textbf{Overview of the skin lesion segmentation task.} We evaluate the performance of \derm against nine pretrained models on the skin lesion segmentation task across five public datasets. The comparison is illustrated in two subfigures. (a) The model in the first subfigure uses a linear head for segmentation, while (b) the second subfigure uses an UpperNet segmentation head. The performance is reported using two widely adopted segmentation metrics: DICE coefficient and Jaccard Index (JAC).}
    \label{fig:seg_result_compare_foundation}
\end{extendeddatafigure*}

\begin{extendeddatafigure*}[htbp!]
    \centering
    \includegraphics[width=1\linewidth]{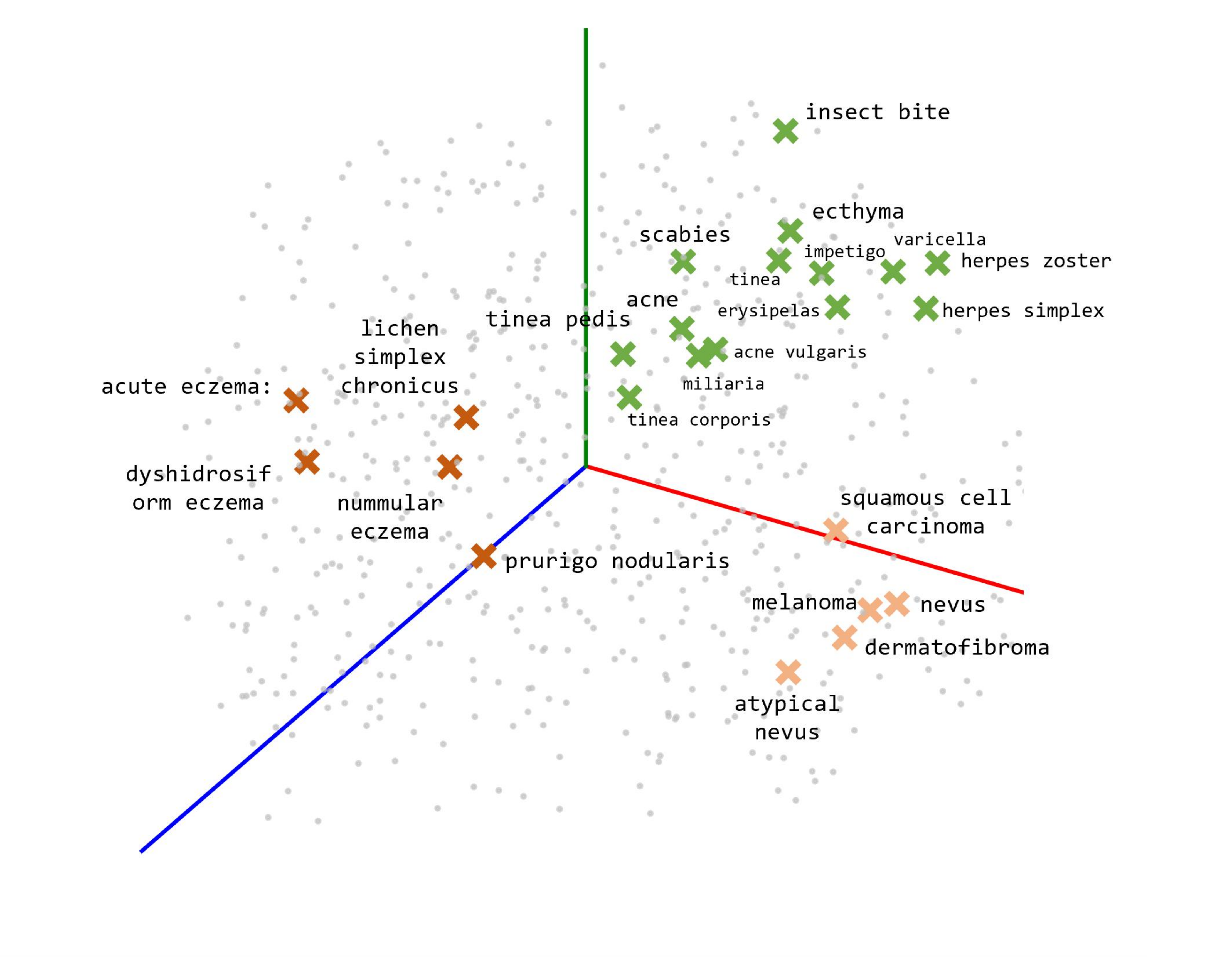}
    \caption{\textbf{t-SNE Analysis of Disease Embeddings (Pretraining Stage).} Before pretraining, we input disease descriptions into the large language model \coder and extract the corresponding CLS token embeddings. To explore the semantic structure already captured by the model, we project these embeddings into a multi-dimensional space using t-SNE, enabling visualization of inter-disease relationships.
}
    \label{fig:embedding}
\end{extendeddatafigure*}

\begin{extendeddatafigure*}[htbp!]
    \centering
    \includegraphics[width=1\linewidth]{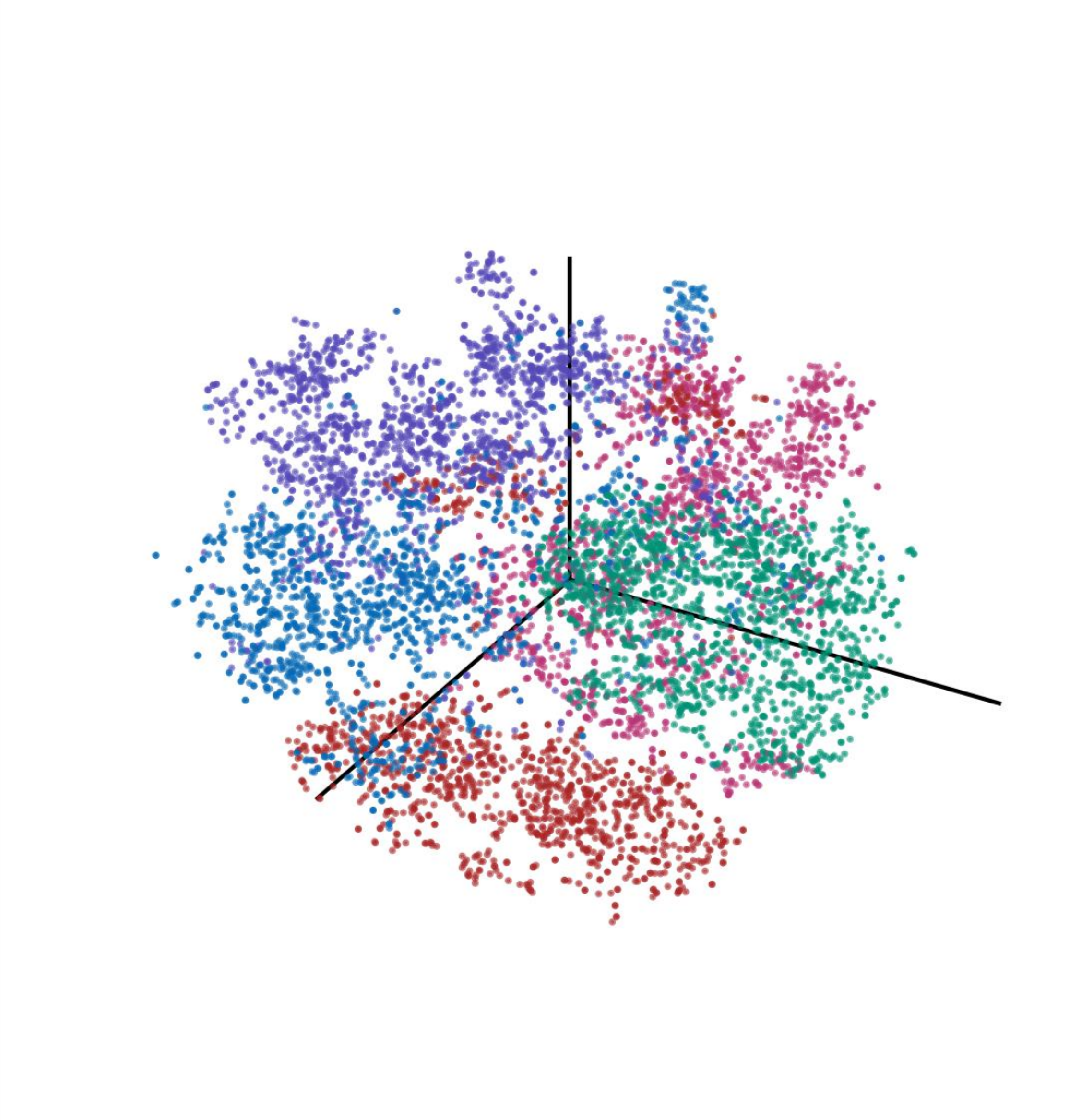}
    \caption{\textbf{t-SNE Analysis of Disease Embeddings (Pretraining Model Output).} We input images from the \acnezerofour and DermNet datasets into our pretrained model and extract the corresponding CLS token embeddings. These embeddings are then projected into a multi-dimensional space using t-SNE, allowing visualization of the learned semantic relationships between diseases.} 
    \label{fig:image_tsne}
\end{extendeddatafigure*}

\clearpage

\begin{supplementarydatatable*}[t]
\centering
\small
\caption{\textbf{Summary of Model Architectures, Parameters Counts, and Types.}}
\label{tab:model_summary}
\resizebox{\textwidth}{!}{
\begin{tabular}{llll}
\toprule
\textbf{Models} & \textbf{Vision model architecture} & \textbf{Parameters} & \textbf{Model type} \\
\midrule
 \resnet     & Resnext101\_32x8d & 86,742,336   & \multirow{2}{*}{General Vision Supervised Models}  \\
 \vit        &  \vit      & 87,002,880   & \\
 \cdashline{1-4}[0.5pt/2pt]
 \siglip     &  \vit      & 87,001,344   & \multirow{2}{*}{General Vision-Language Models} \\
 \clip       &  \vit      & 87,395,328   & \\
 \cdashline{1-4}[0.5pt/2pt]
 \biomedclip &  \vit      & 87,002,880   & \multirow{2}{*}{Medical Vision-Language Models} \\
 \medimage   & DaViT-Large       & 360,632,320  & \\
 \cdashline{1-4}[0.5pt/2pt]
 \dino       &  \vit      & 87,021,312   & General Self-supervised Model \\
 \cdashline{1-4}[0.5pt/2pt]
 \lvm     &  \vit      & 85,737,728   & \multirow{2}{*}{Medical Self-supervised Model} \\
 \panderm    & VIT-Large         & 304,931,840  & \\
 \cdashline{1-4}[0.5pt/2pt]
 \derm       &  \vit      & 87,021,312   & Medical Hybrid Supervised Model \\
\bottomrule
\end{tabular}}
\end{supplementarydatatable*}

\begin{supplementarydatatable}[t]
\centering
\small
\caption{\textbf{Physician Diagnostic Accuracy by Clinical Experience Categories.} The table summarizes the work experience of 23 physicians, grouped into three categories based on years of clinical experience. For each group, diagnostic accuracy is reported under three conditions: aggregated across both datasets, and separately on each dataset.}
\begin{tabular}{@{} c c c c c c @{}}
\hline
\textbf{Doctor} & \textbf{Work Experience} & \textbf{Experience Category} & \textbf{ALL} & \textbf{\classtwo} & \textbf{\classfive} \\
\hline
CRR & 0 year  & \(<3\) & \multirow{10}{*}{67.82} & \multirow{10}{*}{60.75} & \multirow{10}{*}{82.31} \\
CLL & 1 year  & \(<3\) &  &  &  \\
XZN & 1 year  & \(<3\) &  &  &  \\
HJH & 1 year  & \(<3\) &  &  &  \\
TZY & 1 year  & \(<3\) &  &  &  \\
XQY & 2 year  & \(<3\) &  &  &  \\
OYHF & 2 year & \(<3\) &  &  &  \\
ZYJ & 2 year  & \(<3\) &  &  &  \\
ZZ & 2 year   & \(<3\) &  &  &  \\
JZY & 2 year  & \(<3\) &  &  &  \\
 \cdashline{1-6}[0.5pt/2pt]
LMM & 3 year  & \(\geq 3\) & \multirow{9}{*}{75.26} & \multirow{9}{*}{72.64} & \multirow{9}{*}{80.63} \\
LY  & 3 year  & \(\geq 3\) &  &  &  \\
JLF & 3 year  & \(\geq 3\) &  &  &  \\
HY  & 3 year  & \(\geq 3\) &  &  &  \\
YY  & 3 year  & \(\geq 3\) &  &  &  \\
ZXZ & 3 year  & \(\geq 3\) &  &  &  \\
LA  & 4 year  & \(\geq 3\) &  &  &  \\
YJG & 4 year  & \(\geq 3\) &  &  &  \\
LCX & 4 year  & \(\geq 3\) &  &  &  \\
 \cdashline{1-6}[0.5pt/2pt]
FWM & 5 year  & \(\geq 5\) & \multirow{3}{*}{84.66} & \multirow{3}{*}{86.56} & \multirow{3}{*}{80.77} \\
NXL & 5 year  & \(\geq 5\) &  &  &  \\
WWJ & 7 year  & \(\geq 5\) &  &  &  \\
SX  & 8 year  & \(\geq 5\) &  &  &  \\
\hline
\end{tabular}
\end{supplementarydatatable}

\begin{supplementarydatatable*}[h!]
\centering
\footnotesize
\begin{tabular}{|>{\centering\arraybackslash}m{2.5cm}|p{5.5cm}|p{5.5cm}|}
\hline

 &
\makecell[c]{\includegraphics[width=5cm,height=3cm]{figure/caption_vis/110.pdf}} &
\makecell[c]{\includegraphics[width=5cm,height=3cm]{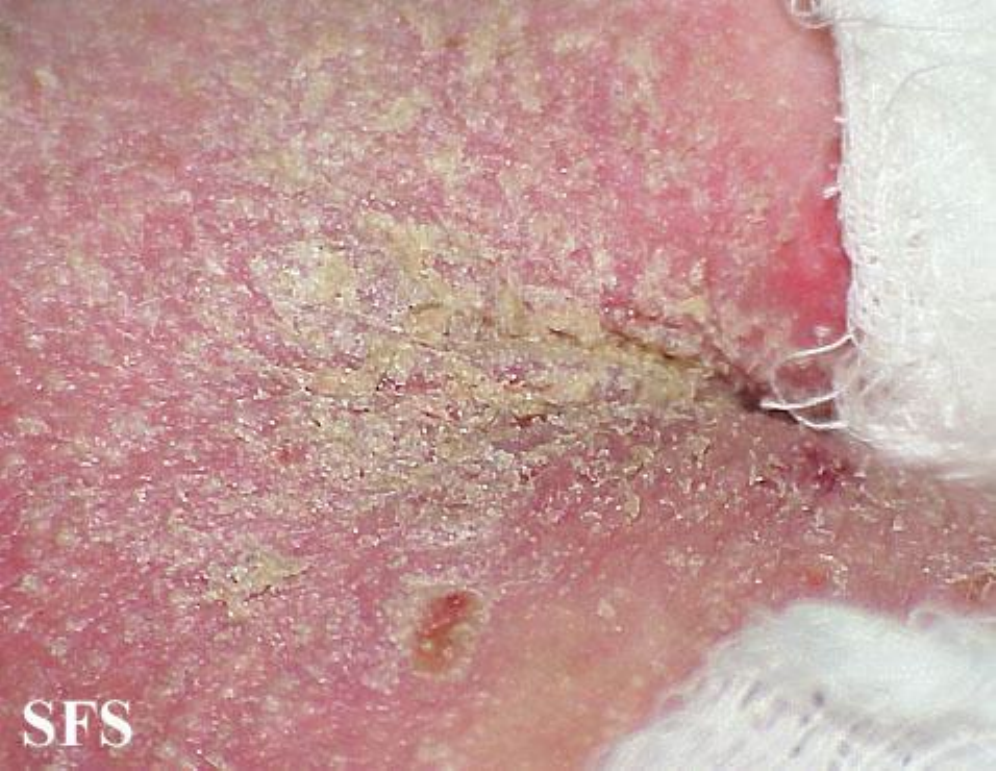}} \\
\hline

\textbf{Ground Truth} &
The patient presents with rectangular gray-black patches on the skin, showing signs of hyperkeratosis on the surface, raising the clinical suspicion of seborrheic keratosis. This condition is typically a chronic, recurrent skin disease characterized by excessive epidermal keratinization and accumulation of keratin, commonly found on the face, scalp, and other areas. &
Patients present with symptoms such as erythema, erosion, exudation, and sinus tracts on the skin, which lead to a diagnosis of scabies based on clinical features. Scabies is a common skin infestation caused by the mite Sarcoptes scabiei, characterized by symptoms including itching, rash, erythema, and erosion.
 \\
\hline

\textbf{\derm} &
The patient presents with rectangular gray-black patches on the skin, showing signs of hyperkeratosis on the surface, raising the clinical suspicion of seborrheic keratosis. This condition is typically a chronic, recurrent skin disease characterized by excessive epidermal keratinization and accumulation of keratin, commonly found on the face, scalp, and other areas. &
The photo shows redness, scales, and crusts on the skin, consistent with the clinical presentation of scabies. Scabies is a contagious skin disease caused by the infestation of the scabies mite, commonly seen in areas with poor hygiene, overcrowding, and low socioeconomic status. \\
\hline

\textbf{\medimage} &
The patient presents with black-brown papules with a smooth surface, slightly raised edges, and a central depression, consistent with the clinical features of seborrheic keratosis. Seborrheic keratosis is a common skin condition characterized by localized overgrowth of keratinocytes, leading to the formation of papules or plaques on the skin. This condition is usually benign but may require further evaluation and treatment by a dermatologist. &
This is a photo of the trunk showing large red patches with central atrophy and erosion, consistent with the clinical presentation of eczema. Eczema is a common chronic inflammatory skin condition that can be caused by various factors such as allergies, infections, or autoimmune reactions. Treatment typically involves topical medications and avoiding triggering factors. \\
\hline

\textbf{\dino} &
The photo shows a dark brown patch with uneven color, irregular shape, and clear borders, consistent with the clinical presentation of melanoma. Melanoma is a malignant skin tumor that typically presents as a dark brown patch with irregular shape, uneven color, and clear borders. It is recommended to seek medical attention promptly for further evaluation and treatment. &
This is a photo of a skin lesion on the ear, showing a large area of erythema with clear borders, consistent with the clinical presentation of eczema. Eczema is a common inflammatory skin condition, often caused by allergic reactions or dry skin, characterized by symptoms such as redness, itching, and scaling. Treatment typically involves topical medications and moisturizing care. \\
\hline

\textbf{\panderm} &
The patient presents with black papules on the trunk, with a rough surface and a darker color around them, which are consistent with the clinical presentation of seborrheic keratosis. Seborrheic keratosis is a common skin condition characterized by localized overgrowth of keratin, forming rough, darker-colored papules. This condition is usually benign, but in some cases, it may require treatment. &
This is a photo of the trunk showing a large ulcer with a black crust on the surface, which is consistent with the clinical presentation of squamous cell carcinoma. Squamous cell carcinoma is a common malignant tumor that typically originates from the squamous cells of the skin. It can occur in any part of the body, but is more common in sun-exposed areas such as the head and neck. Early symptoms may include localized skin changes, ulcers, and crusting, and in severe cases, it can lead to bleeding and pain. \\
\hline

\end{tabular}
\caption{Image captioning examples generated by different models.}
\end{supplementarydatatable*}

\begin{supplementarydatatable*}[h!]
\centering
\footnotesize
\begin{tabular}{|>{\centering\arraybackslash}m{2.5cm}|p{5.5cm}|p{5.5cm}|}
\hline

 &
\makecell[c]{\includegraphics[width=5cm,height=3cm]{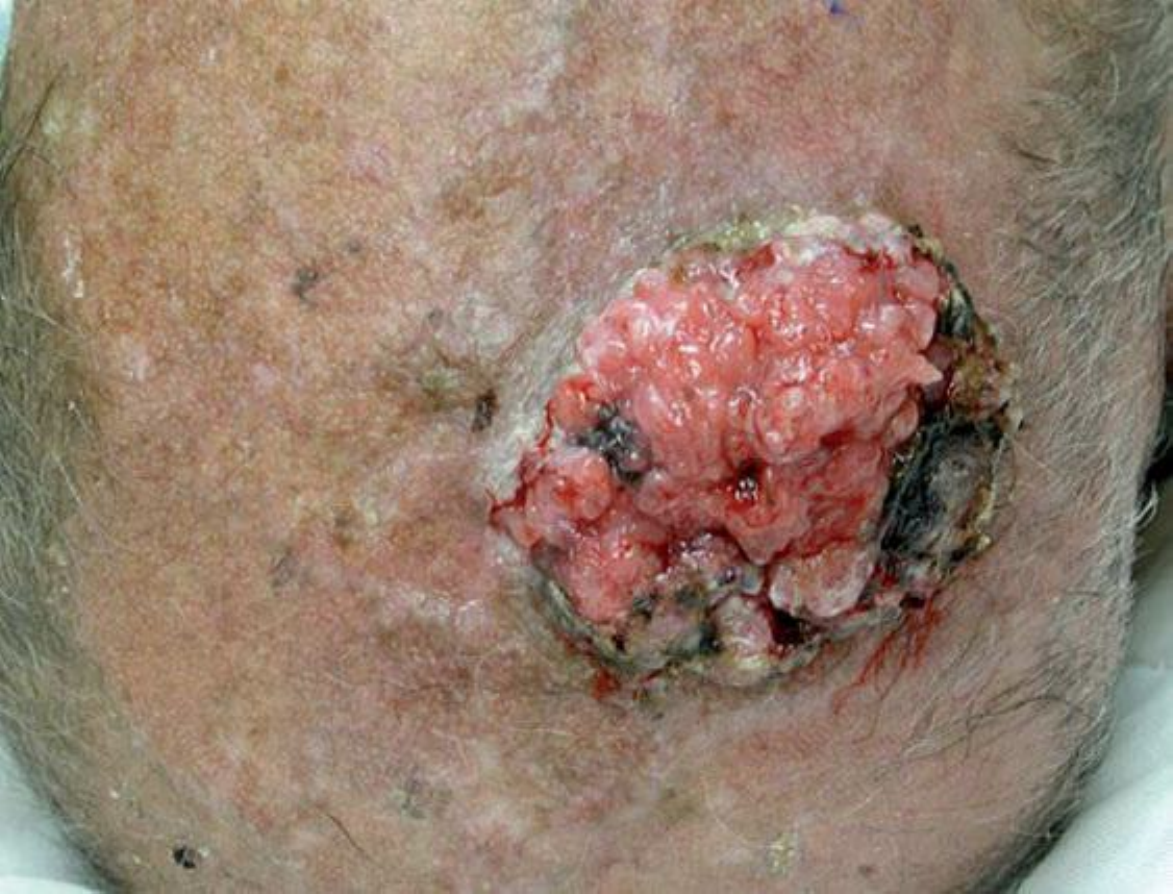}} &
\makecell[c]{\includegraphics[width=5cm,height=3cm]{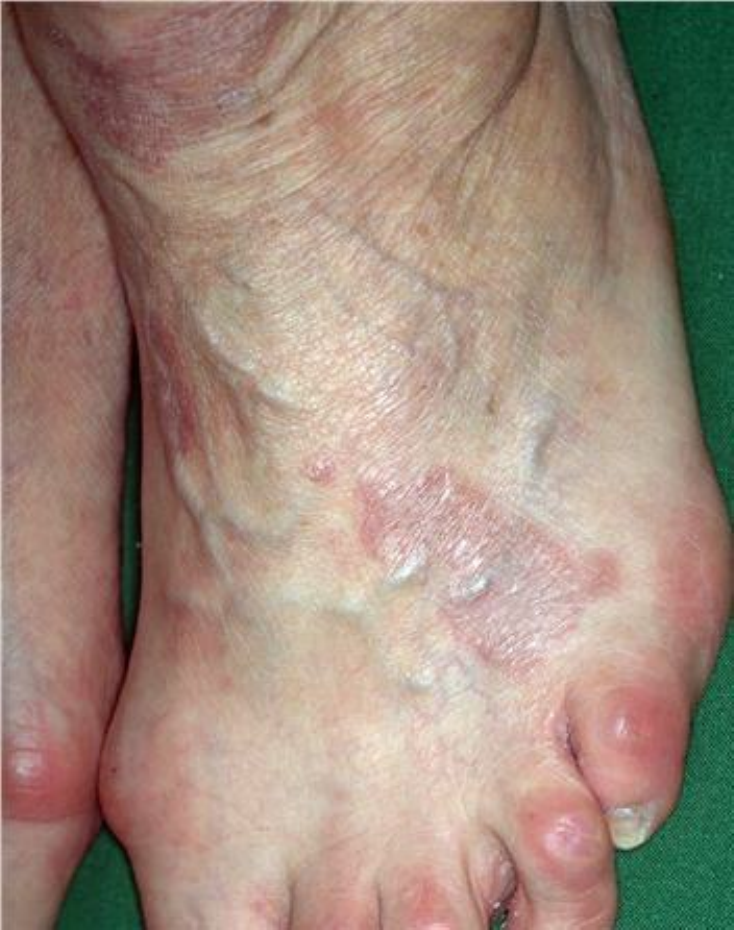}} \\
\hline

\textbf{Ground Truth} &
The patient presents with a round ulcer with granulation tissue at the base, accompanied by bleeding and scabbing. According to the pathological findings, the diagnosis is squamous cell carcinoma. &
The patient had flaky erythema on the dorsum of his feet, with central regression and slightly raised edges. It is consistent with the performance of granuloma annulare, and dermatopathological examination is recommended for further diagnosis.
 \\
\hline

\textbf{\derm} &
This is a photo of a skin lesion on the head, showing a large ulcer with a central crust and raised edges. Based on the characteristics of the lesion, the preliminary diagnosis is squamous cell carcinoma. Squamous cell carcinoma is a common type of skin cancer that typically originates from the squamous cells in the epidermis. It is recommended to seek medical attention promptly for further diagnosis and treatment. &
This is a photo of the foot showing multiple dark red patches with clear borders, consistent with the clinical presentation of granuloma annulare. Granuloma annulare is a common skin condition characterized by localized skin lesions, often seen on the feet and hands. The exact cause is not clear, but it may be related to factors such as genetics and immune system abnormalities. Treatment typically involves topical medications and local care. \\
\hline

\textbf{\medimage} &
This is a photo of the trunk showing a large ulcer with raised edges, central ulceration, and crusting, consistent with the clinical presentation of squamous cell carcinoma. Squamous cell carcinoma is a common type of skin cancer that typically originates from the squamous cells in the epidermis. Early symptoms may include ulcers, crusting, and raised edges, and further investigation is needed for confirmation. &
This is a photo of the foot showing multiple red papules on the dorsum of the foot, some of which have merged into plaques, consistent with the clinical presentation of granuloma annulare. Granuloma annulare is a chronic inflammatory skin condition characterized by red or purple papules or plaques on the skin, often accompanied by itching or pain. The exact cause of this condition is not clear, but it may be related to factors such as immune system abnormalities. Treatment options include topical medications, oral medications, and laser therapy. \\
\hline

\textbf{\dino} &
This is a photo of a skin lesion on the scalp, showing a large ulcer with a central crust, raised edges, and surrounding erythema, consistent with the clinical presentation of squamous cell carcinoma. Squamous cell carcinoma is a common type of skin cancer that typically originates from the epidermis, commonly found on the head, neck, and other areas. Treatment options include surgical excision, radiation therapy, and chemotherapy. &
This is a photo of a foot showing multiple red patches with scales on the dorsum of the foot, consistent with the clinical presentation of psoriasis. Psoriasis is a chronic inflammatory skin condition characterized by red patches, scales, and itching, commonly found on the scalp, trunk, and limbs. Treatment options include topical medications, phototherapy, and oral medications. \\
\hline

\textbf{\panderm} &
This is a photo of a skin lesion on the lower limb, showing a large ulcer with a black base and a scab on the surface. Based on the characteristics of the lesion, the diagnosis is squamous cell carcinoma. Squamous cell carcinoma is a common type of skin cancer that typically originates from squamous cells in the epidermis. Early symptoms may include skin thickening, ulcers, and scabs, and prompt medical attention and treatment are recommended. &
This is a photo of the hand showing multiple red patches with scales on the dorsum of the hand, consistent with the clinical presentation of psoriasis. Psoriasis is a chronic inflammatory skin disease characterized by red patches, scales, and itching, commonly affecting joints and skin. Treatment options include topical medications, phototherapy, and systemic medications. \\
\hline

\end{tabular}
\caption{Image captioning examples generated by different models.}
\end{supplementarydatatable*}

\begin{supplementarydatafigure*}[htbp!]
    \centering
    \includegraphics[width=1\linewidth]{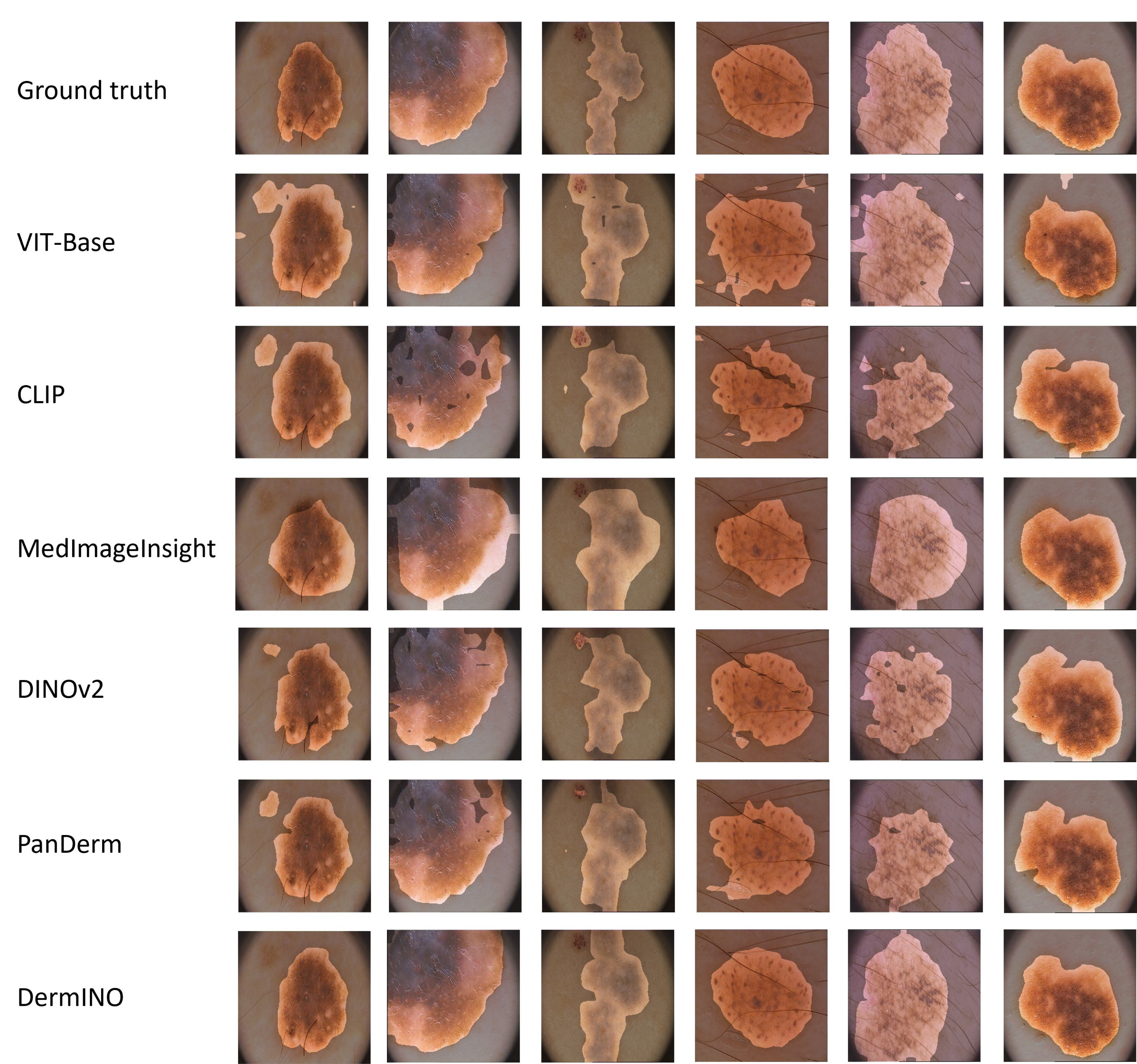}
    \caption{\textbf{Qualitative results of skin lesion segmentation.} The first row presents the ground truth masks, followed by the predicted segmentation results from \derm and six other representative models.
}
    \label{fig:segment_visualization}
\end{supplementarydatafigure*}


\end{document}